\documentclass[10pt,journal,compsoc]{IEEEtran}

\usepackage{graphicx}
\usepackage{multirow}
\usepackage{hhline}
\usepackage{makecell}
\usepackage[draft]{changes}
\usepackage{pifont}
\usepackage{amssymb}
\usepackage{orcidlink}
\newcommand{\cmark}{\ding{51}}%
\newcommand{\xmark}{\ding{55}}%

\newcount\commentTon
\newcommand{\vise}[1]{\ifnum\commentTon=0{\color{orange} \it [VISE: #1]}\fi} 
\newcommand{\rossi}[1]{\ifnum\commentTon=0{\color{red} \it [ROSSI: #1]}\fi} 
\newcommand{\ken}[1]{\ifnum\commentTon=0{\color{brown} \it [KEN: #1]}\fi} 
\newcommand{\steve}[1]{\ifnum\commentTon=0{\color{blue} \it [STEVE: #1]}\fi} 
\newcommand{\diego}[1]{\ifnum\commentTon=0{\color{purple} \it [DIEGO: #1]}\fi} 
\def\BibTeX{{\rm B\kern-.05em{\sc i\kern-.025em b}\kern-.08em
    T\kern-.1667em\lower.7ex\hbox{E}\kern-.125emX}}

\ifCLASSOPTIONcompsoc
  \usepackage[nocompress]{cite}
\else
  \usepackage{cite}
\fi
\usepackage[numbers]{natbib}
\usepackage{graphicx}
\usepackage{epstopdf}
\ifCLASSINFOpdf
\else
\fi
\usepackage{xurl}

\begin{document}
%
\title{Forecasting Workload in Cloud Computing: Towards Uncertainty-Aware Predictions and Transfer Learning}
%
%
%
%

\author{Andrea~Rossi{\orcidlink{0000-0002-0780-2810}},
        Andrea~Visentin{\orcidlink{0000-0003-3702-4826}},
        ~Diego~Carraro{\orcidlink{0000-0002-2857-0473}},
        ~Steven~Prestwich{\orcidlink{0000-0002-6218-9158}},
        ~and~Kenneth~N.~Brown{\orcidlink{0000-0003-1853-0723}}
\IEEEcompsocitemizethanks{\IEEEcompsocthanksitem A. Rossi is with the Centre for Research Training in Artificial Intelligence, University College Cork, Ireland.\protect\\
E-mail: a.rossi@cs.ucc.ie
\IEEEcompsocthanksitem A. Visentin, D. Carraro, S. Prestwich and K. N. Brown are with the Insight Centre for Data Analytics, School of Computer Science, University College Cork, Ireland.\protect\\
E-mail: a.visentin@ucc.ie, diego.carraro@insight-centre.org, \{s.prestwich, k.brown\}@cs.ucc.ie}
 }

\IEEEtitleabstractindextext{%
\begin{abstract}

Predicting future resource demand in Cloud Computing is essential for optimizing the trade-off between serving customers' requests efficiently and minimizing the provisioning cost. Modelling prediction uncertainty is also desirable to better inform the resource decision-making process, but research in this field is under-investigated. In this paper, we propose univariate and bivariate Bayesian deep learning models that provide predictions of future workload demand and its uncertainty. We run extensive experiments on Google and Alibaba clusters, where we first train our models with datasets from different cloud providers and compare them with LSTM-based baselines. Results show that modelling the uncertainty of predictions has a positive impact on performance, especially on service level metrics, because uncertainty quantification can be tailored to desired target service levels that are critical in cloud applications. Moreover, we investigate whether our models benefit transfer learning capabilities across different domains, i.e.\ dataset distributions. Experiments on the same workload datasets reveal that acceptable transfer learning performance can be achieved within the same provider (because distributions are more similar). Also, domain knowledge does not transfer when the source and target domains are very different (e.g.\ from different providers), but this performance degradation can be mitigated by increasing the training set size of the source domain. 


\end{abstract}

}

\maketitle

\IEEEdisplaynontitleabstractindextext

%
\IEEEpeerreviewmaketitle

\IEEEraisesectionheading{\section{Introduction}\label{sec:introduction}}

\begin{figure*}[htbp]
\centerline{\includegraphics[width=0.8\linewidth]{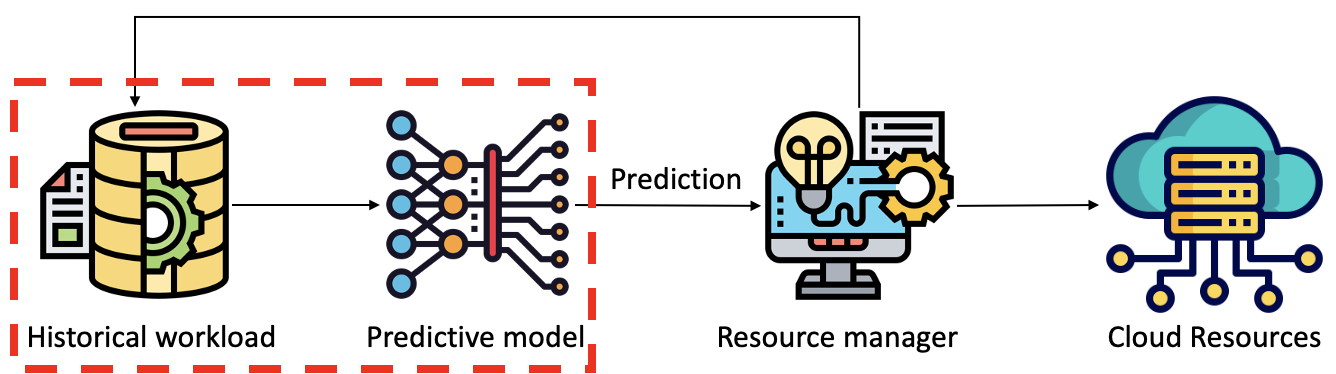}}
\caption{The workload prediction scheme in cloud computing. Our work focuses on building and evaluating future workload predictions (i.e.\ red box).}
\label{fig:prediction_scheme}
\end{figure*}

\IEEEPARstart{C}{loud} computing services have gained enormous popularity in the last few years because they reduce the cost of business and increase the productivity of industries \cite{muller2015benefits}. It is expected that cloud computing demand will rise steadily in the future, especially due to the recent advances in Artificial Intelligence (AI) and Big Data, whose workloads require massive computational resources \cite{marr2020tech}. To maintain their business sustainability and cost-effectiveness, cloud providers are required to efficiently optimize management and dynamic allocation of computing resources, such as processing power, memory and storage, based on the current and future workload demand. Accurately predicting cloud workload is, therefore, an essential precondition to meet such optimization goals. Typically, the workload prediction scheme is a loop (see Figure \ref{fig:prediction_scheme}), where the forecast is provided by a predictive model, which is trained on the historical workload data. Based on such predictions, cloud managers (or automated algorithms) can, for example, implement resource allocation policies and elastic scaling or perform workload balancing across resources. Over time, the workload history evolves and feeds back as newly available data to update the predictive model.

Workload forecasting is a challenging task due to the dynamic of workload patterns, which exhibit spikes and irregularities \cite{tirmazi2020borg}. Research tackles the problem as a time series analysis task and has proposed statistical methods such as ARIMA, Machine Learning (ML) approaches such as linear regression and decision trees and recently, Deep Learning (DL) architectures such as Long Short-Term Memory (LSTM). The majority of these models provide point forecast predictions of the future workload, but they do not consider the uncertainty of such predictions.
However, predicting probability distribution instead allows one to account for the uncertainty of the model due to the noise in the data (i.e.\ aleatory uncertainty) and due to the lack of representativeness of the training data (i.e.\ epistemic uncertainty). Moreover, it can be leveraged by cloud managers to make more informed decisions when optimizing resource allocation, approaching the problem based on the degree of a prediction’s confidence. For example, when the uncertainty is high, the safety margin resources allocated are increased compared to when the models have more confidence in the predictions. The literature has proposed uncertainty-aware solutions for workload prediction in cloud computing systems \cite{rossi2022bayesian, minarolli2014cross, minarolli2017tackling}, but the domain is still under-investigated and focuses mostly on modelling only the aleatoric uncertainty.


Moreover, the literature does not assess whether the proposed predictors \textit{generalize} well to out-of-distribution data. In ML, this problem is generally tackled with Transfer Learning (TL) \cite{weiss2016surveytransfer}, where a model is trained on one or several source domain(s) (i.e.\ in-distribution datasets) and is used or adapted to predict a different but related target domain (i.e.\ out-of-distribution dataset)\footnote{In the literature, this is sometimes referred to as Domain Adaptation.}. In the cloud computing scenario, transfer learning is useful, for example, when a cloud provider establishes a new data centre in a previously untapped region, and therefore, limited or no historical workload data is available to train a workload predictor for such region. In this case, one possibility to obtain a workload forecast for the new region is to leverage a predictor trained on data from other regions, i.e.\ one that can transfer its predictive capabilities to the new region.

In this paper, we conduct a comprehensive investigation into models that provide uncertainty-aware predictions for workload in cloud computing, and we test their transfer learning capabilities. We build such investigation on our previous work \cite{rossi2022bayesian}, where we proposed a Bayesian Neural Network (i.e.\ HBNN) and a Probabilistic LSTM (i.e.\ LSTMD). The former architecture captures the epistemic and aleatoric uncertainty of the prediction, while the latter captures only the aleatoric uncertainty. The contributions and the findings of our work are summarised as follow:





\begin{itemize}
    \item We compare models that provide univariate and bivariate uncertainty-aware predictions of processing unit workloads in terms of CPU/GPU and memory usage. 
    
    \item We modify the HBNN and LSTMD architectures in \cite{rossi2022bayesian} by adding additional layers to improve their representation power to fit the increasing training data and the bivariate labels. The goal is to create and hyperparametrise more complex networks with improved domain generalization capabilities. As a baseline, we employ a deterministic point-estimate LSTM, i.e.\ one that does not provide uncertainty for the predictions.

    \item We preprocess twelve widely-used workload traces from Google Cloud 2011 \cite{reiss2011google} and 2019 \cite{clusterdata:Wilkes2020a} and Alibaba Cluster 2018 \cite{jiang2020characterizing} and 2020 \cite{weng2022mlaas}, and we open source the resulting datasets\footnote{\url{https://github.com/andreareds/TowardsUncertaintyAwareWorkloadPrediction}}. We make this procedure detailed and explicit for reproducibility as, to the best of our knowledge, the current literature is inconsistent and lacking in detail. 

    \item We design and run experiments composed of two parts where we assess the effectiveness of our uncertainty-aware models on the prediction task at hand using the preprocessed datasets. We evaluate their runtime performance and their prediction quality in terms of a wide range of accuracy and service level metrics. 

    \item The first part of the experiments compares univariate/bivariate models trained on one or multiple datasets. Our findings show that HBNN and LSTMD offer a one-model-fits-all solution that provides accurate bivariate predictions with uncertainty quantification on multiple datasets. 

    \item The second part of the experiments assesses the transfer learning capabilities of the bivariate HBNN trained on multiple datasets. We design and run several prediction scenarios where we apply zero-shot or fine-tuning on a pre-trained model, each testing the model's performance using a different combination of source and target domains (datasets). Results show that the model has acceptable transfer learning capabilities when trained with a substantial amount of data or when source and target domains are similar (from the same providers), while struggles if the training set is small or the source and target domains are from different cloud providers. 

\end{itemize}

The remainder of the paper is structured as follows. Section \ref{sec:RelatedWork} survey the related work. We describe the proposed forecasting models and the baseline in Section \ref{sec:predmodels}, and Section \ref{sec:experiments} presents the experimental framework. Section \ref{sec:results} reports the results of the experiments, and Section \ref{sec:conclusion} outlines the conclusions and future works.

\section{Related Work}\label{sec:RelatedWork}

In cloud computing management, workload prediction plays an important role that has been broadly studied over the past twenty years. In this section, we outline relevant literature on statistical, ML and DL forecasting approaches and the role of uncertainty and transfer learning.

\subsection{Workload forecasting}
Researchers and practitioners have tackled forecasting workloads with statistics-based at first, then traditional ML-based, and finally transitioning to DL-based, considered state-of-the-art in most modern cloud applications. 

Statistics-based approaches employ mathematical models to identify repetitive patterns and trends in the data, but they often struggle to capture the high complexity and variability of cloud workloads accurately. ARIMA-based \cite{hyndman2018forecasting} models are simple yet powerful statistical techniques that essentially combine three components: the AutoRegressive (AR) component ensures predictions are linear combinations of historical values of the target variable; the Integrated (I) component helps make the series stationary through differencing operations; and the Moving Average (MA) component leverages past forecast errors to make predictions. ARIMA models are effective at capturing mainly stationary trends and seasonality effectively. They have been used, for example, for predicting the future host workload in distributed systems \cite{dinda2000host} and for predicting HTTP requests \cite{calheiros2014workload}. ARIMA is now a popular baseline approach for evaluation purposes, e.g.\ \cite{vazquez2015time, baldan2016forecasting, kumar2021performance}, as more advanced approaches and prediction tasks arose in recent years. Other statistical techniques have been applied to workload forecasting. For example, exponential smoothing \cite{brown1956exponential}  and GARCH \cite{bollerslev1986generalized} are applied as baselines to forecast CPU/memory demand (e.g.\ \cite{di2012host, yang2014new, di2014google, wang2021predicting} and \cite{barati2015hybrid}, respectively) for autoscaling applications. The former models leverage an autoregressive component to capture simple patterns and trends, while the latter leverage autoregressive and moving average components and are designed to model and forecast time series where variance changes over time, i.e.\ heteroskedasticity.

ML-based models have been shown to outperform statistical methods and improve prediction accuracy \cite{gao2020machine}. Indeed, they offer numerous benefits over traditional statistical methods because they can handle complex and diverse data more efficiently and successfully capture non-linear patterns in the data. On the other hand, they require manual feature extraction and hyperparameter tuning, which is sometimes expensive. Examples in the literature include logistic regression \cite{ajila2013cloud}, random forest \cite{shishira2021novel}, K-nearest neighbours regression \cite{didona2015enhancing}, support vector regression \cite{zhong2018load, gao2020machine} and ensemble methods \cite{banerjee2021efficient}.

Although training DL-based architectures is expensive in terms of data availability and computational resources, such models are now considered state-of-the-art because they outperform both statistical and ML-based alternatives in most of the cloud applications. In particular, sequential models have become a popular choice for their inner time-sensitive nature, effective at capturing long-term dependencies and dynamicity of workload data. One such model is the Recurrent Neural Network (e.g.\ used in \cite{zhang2016workload, duggan2017predicting}), and another is its advanced versions, i.e.\ Long Short Term Memory (e.g.\ used in \cite{bi2021integrated, wang2021predicting}) and Gated Recurrent Unit (e.g.\ used in \cite{xu2022esdnn}). In this paper, we investigate an LSTM-based model to perform workload forecasting (see section \ref{sec:predmodels}). 
When dealing with cloud workload time series data, it is common to see intricate patterns over extended periods. To capture these long-range dependencies, attention mechanisms \cite{vaswani2017attention} and transformers are often used. Attention mechanisms help models focus on the relevant parts of the input sequence, while transformers enable parallel data processing and eliminate the need for recurrence. These techniques have revolutionized sequence modelling and applied to the cloud workload forecasting problem \cite{xi2021attention, arbat2022wasserstein, bao2022long}. 

Finally, researchers have explored combining two or more models above into ensembles to enhance forecasting accuracy and model robustness. Examples of this kind are \cite{liu2016quantitative, zia2017adaptive, kim2020forecasting}. However, in this paper, we focus on comparing individual architectures and leave the additional investigation of combining such architectures into ensemble models as future work.

\subsection{Uncertainty-aware forecasting}
The models discussed earlier provide a point prediction of future workload with no indication of confidence in such prediction. This can be problematic because, due to the lack of explainability of the models, these predictions may be either underconfident or overconfident, and we cannot distinguish between them. In critical use cases such as healthcare or autonomous vehicles, incorrect, overconfident predictions can be dangerous. Measuring uncertainty in time series forecasting can be done through various methods \cite{foldesi2022comparison}. One of them is ensemble forecasting, which involves multiple models, each with slightly varying initial conditions, to produce a range of possible outcomes and their probabilities. This method is commonly used in weather forecasting \cite{gneiting2005weather}. Another approach is probabilistic forecasting, which estimates the probability of various future outcomes and the complete set of probabilities represents a probability forecast \cite{foldesi2022comparison}. These uncertainty quantification methods can also be divided into parametric and non-parametric methods. The former does not make strong assumptions about the underlying probability distributions or functional forms of the data. The latter assumes that the data follow a specific distribution, often a normal distribution. In the literature, there is limited investigation of approaches capable of capturing and quantifying the inherent uncertainties in cloud workload data. In ML, uncertainty is often categorised as either epistemic, i.e.\ due to insufficient training data, or aleatoric, i.e.\ caused by stochasticity of observations due to noise and randomness. The design of the approach influences the type of uncertainty that can be modelled \cite{abdar2021review}. 
In the workload prediction task, early attempts have been made to capture such uncertainties through confidence intervals, e.g.\ \cite{arima_calheiros}, and non-parametric methods such as kernel estimation method, e.g.\ \cite{minarolli2017tackling}. Recently, a popular non-parametric method called Quantile Regression has been applied to workload prediction to model aleatoric uncertainty \cite{mammen2023cuff} through quantile distributions. However, interpreting uncertainty estimates in non-parametric models can be more difficult, as they combine multiple sources of uncertainty \cite{kong2020sde}. In particular, none of these methods aims at quantifying epistemic and aleatoric uncertainties together, and none of them provides bivariate forecasting. To the best of our knowledge, in our previous work \cite{rossi2022bayesian}, we are the first to explore state-of-the-art DL models designed to model both uncertainties under a Bayesian framework. In this paper, we extend this work and provide a more comprehensive investigation of the impact of uncertainty in workload prediction.

\subsection{Transfer Learning}
Transfer Learning (TL) \cite{weiss2016surveytransfer} is a broad machine learning paradigm where a model is pre-trained on a source task or domain and then adapted or fine-tuned to a related target task or domain. Transferring knowledge across tasks is generally known as \textit{task transfer}. Transferring knowledge across domains is generally known as \textit{domain adaptation}, \textit{cross-domain} or \textit{domain generalization} \cite{dou2019domain, wang2022generalizing}, very closely related to each other, so that sometimes the literature is inconsistent in its terminology and does not distinguish between them. In this paper, our work is about transferring knowledge across domains and overlaps the three areas so that we use the general transfer learning umbrella terminology. In particular, we investigate when pre-training is performed on \textit{multiple} source domains to learn domain-invariant representations that can be exploited on a different target domain. Generalization can be achieved by fine-tuning (FT) on the target dataset or in a zero-shot fashion. 

TL has been widely applied in the context of time series classification \cite{fawaz2018transfer} and forecasting \cite{ye2018novel}, showing the effectiveness of this approach, including QoS aspects \cite{hao2021transfer} for workload prediction \cite{liu2022tr}. TL has also been applied in other cloud computing contexts, such as runtime performance prediction \cite{li2020cross, nguyen2012deep} and autoscaling \cite{zhang2021autrascale}. To the best of our knowledge, we are the first to explore transfer learning across domains for workload predictions.

\section{Forecasting Models}\label{sec:predmodels}
\begin{figure}[htbp]
\centerline{\includegraphics[width=\linewidth]{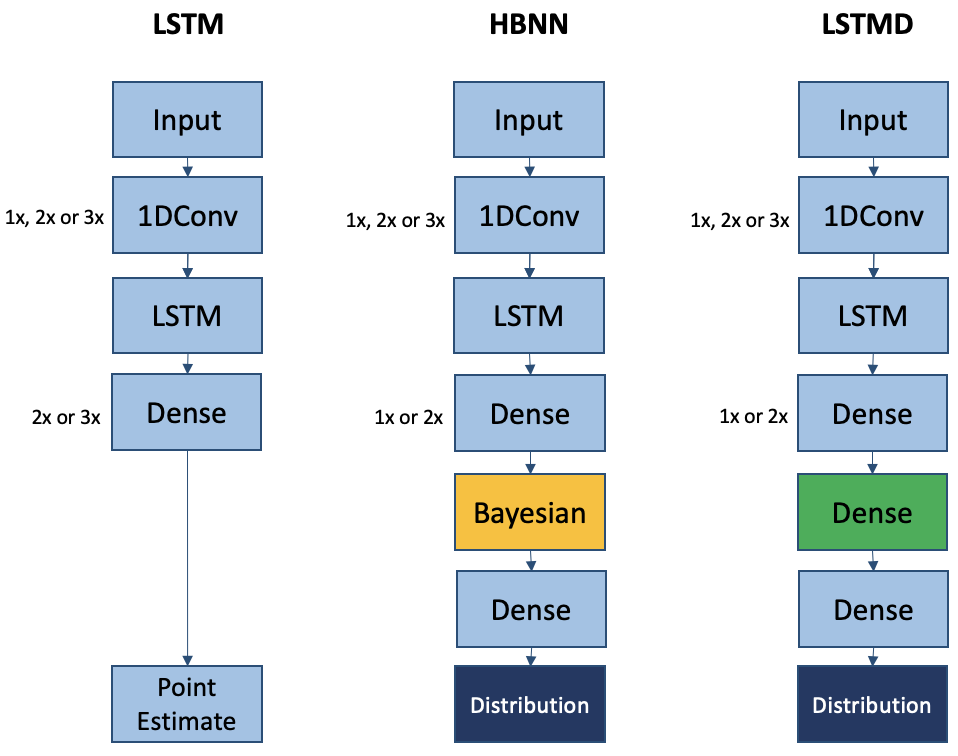}}
\caption{Network architectures of the three models we compare. The presence of the Bayesian layer and the type of output layer characterizes the three architectures.}
\label{fig:architectures}
\end{figure}

This section describes the three forecasting models we compare in our work. A graphical representation of their architectures is depicted in Fig. \ref{fig:architectures}. Hyperparametrization and training details are given in section \ref{sec:experiments}.

\subsection{LSTM}\label{subsec:lstm}

An LSTM-based model that does not provide uncertainty quantification is used as the baseline of our experiments. The network consists of an input layer with a size of 288, corresponding to the past 24 hours of workload. The input size has been found experimentally. The input sequence is filtered by a succession of 1D convolutional (1DConv) layers (between one and three). The convolution is followed by an LSTM layer, widely adopted as effective at learning long-term dependencies and sequential patterns in time-series or sequence data \cite{song2018host}. Combining convolutional and LSTM layers has been proven effective in time series forecasting \cite{leka2021hybrid, patel2022hybrid}. A sequence of dense layers, whose number varies with the training set's size and the prediction type, follows the LSTM layer. We optimize the network weights through the Mean Squared Error (MSE) loss function.

\subsection{HBNN}

Similarly to our previous work \cite{rossi2022bayesian}, we employ a Bayesian Last Layer network to capture epistemic and aleatoric uncertainties. HBNN follows the same architecture of the LSTM-based described in \ref{subsec:lstm}, with the following modifications. A Bayesian layer, designed to model the epistemic uncertainty, is positioned after the first sequence of dense layers. In the Bayesian layer, each weight, instead of being deterministic, has an associated probability distribution that can be trained via variational inference. This means that when the input is propagated through this layer, the distributions of the neurons are sampled, making the output probabilistic. This layer is followed by a dense layer that outputs the parameters (mean and standard deviation) of one or two Gaussian distributions for univariate and bivariate predictions, respectively. This layer has two neurons in the univariate case and four neurons in the bivariate case. The output Gaussian distributions capture the aleatoric uncertainty. Compared to \cite{rossi2022bayesian}, in the experiments of this paper, we perform additional hyperparametrization on the dense layers to account for the greater complexity of training the model on more datasets and acquiring transfer learning capabilities. We optimize the weights through the negative log-likelihood loss function.

\subsection{LSTM Distributional}

The LSTM Distributional (LSTMD) network \cite{rossi2022bayesian} is similar to the HBNN, but a traditional dense layer replaces the Bayesian layer so that epistemic uncertainty is ignored. We again perform the same type of hyperparametrization and use the same loss function for training. 

\section{Experimental framework}\label{sec:experiments}
This section describes the framework we designed for our investigation and how we ran the experiments. 

\subsection{Datasets}
The datasets we use in our experiments are extracted from Google Cloud Trace 2011 and 2019, Alibaba Cluster Trace 2018 and Alibaba Cluster Trace 2020. In the following, we describe such data and how we preprocess it. The resulting twelve datasets can be downloaded from the GitHub repository we share.

\subsubsection{Google Cloud Trace 2011 and 2019}

The Google Cloud Trace 2011 (GC11) \cite{reiss2011google} and 2019 (GC19) \cite{clusterdata:Wilkes2020a} are publicly available datasets collected from the Google Cloud Platform and contain details about the resource utilisation of some cluster cells. In particular, Google Cloud Trace 2011 is composed of 29 days of resource usage collected in May 2011 from 12,500 machines in a single cluster cell; Google Cloud Trace 2019 is composed of data of 29 days from eight different cluster cells accounting for around 10,000 machines each, distributed across different geographical regions around the world. Similarly to other works in literature \cite{kumar2019efficient, liu2016quantitative, herbst2014self}, we preprocess Trace 2011 and Trace 2019 with the following procedure, and for the Trace 2019 we also employ Google BigQuery due to the dataset size, i.e.\ about 2.4TiB compressed. For each cluster, we create a time series dataset that includes the average CPU and average memory usage for all the machines with a 5-minute interval. Missing records are neglected for simplicity. For the program tasks that run only partially in a 5-minute window, we multiply the average resource by a weight corresponding to the fraction of the window in which the task is in execution. Data is finally scaled in the range [0, 1] using a MinMax scaling strategy for speeding up the convergence of the training. We obtain nine datasets of roughly 8k data points per resource (CPU and CPU memory usage).

\subsubsection{Alibaba Cluster Trace 2018 and 2020}
The cluster trace 2018 (AC18) includes the workload history for CPU and memory utilisation of about 4,000 machines in eight consecutive days. The 2020 version (AC20) is a longer trace of around two months and 1,800 machines that contain over 6,500 GPUs. From this trace, we craft two datasets, one related to the CPU and memory usage and one for the GPU and GPU memory usage. We preprocess these traces with the same procedure we used for the Google traces. We obtain one dataset of roughly 2k data points per resource (CPU and CPU memory usage) for Alibaba 2018; and two datasets of roughly 14k data points per resource (GPU, CPU and their memory usage) for Alibaba 2020.

\begin{figure}[htbp]
\centerline{\includegraphics[width=\linewidth]{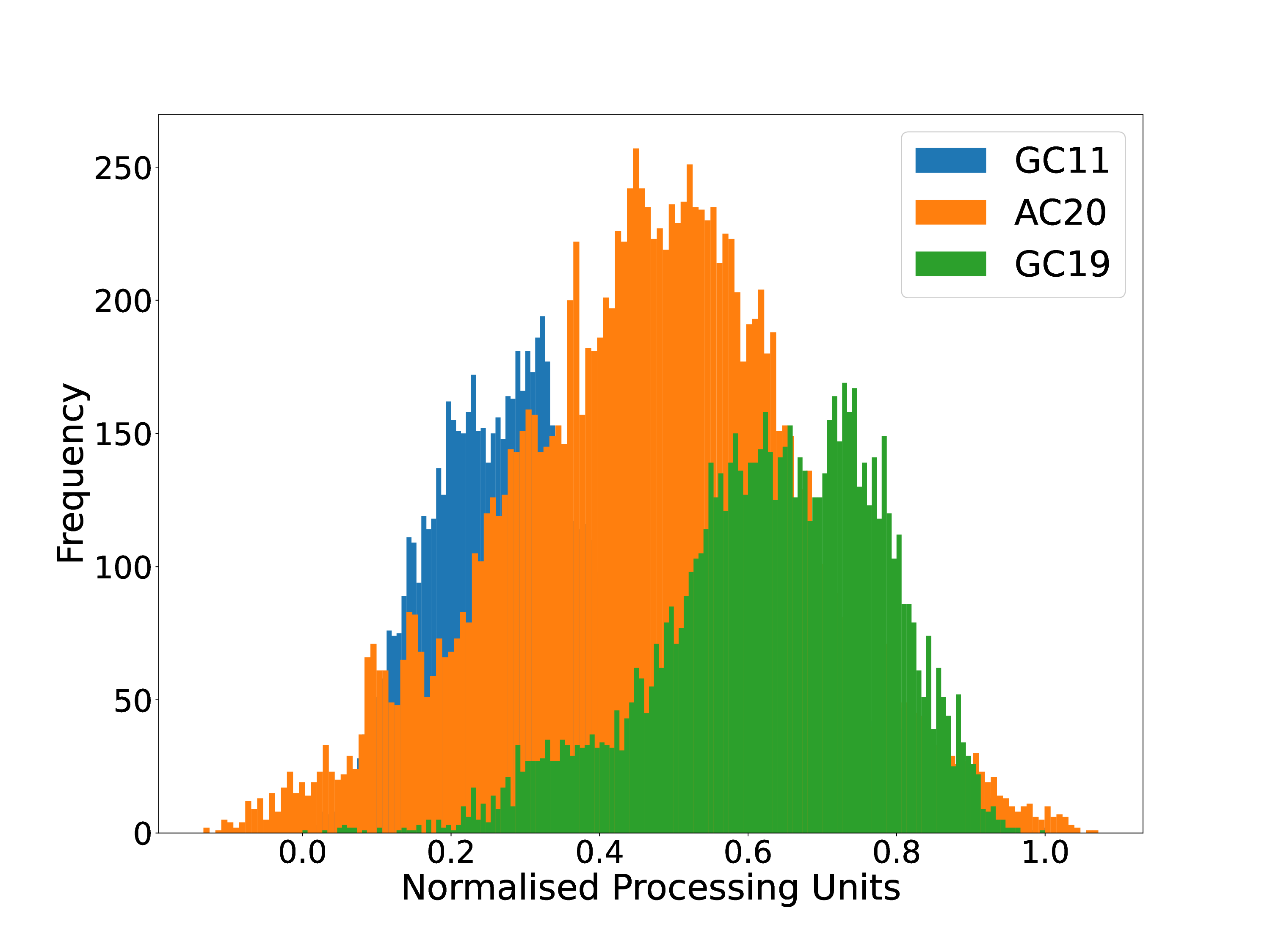}}
\caption{Processing units distributions from different clusters. We plotted only one dataset per cluster, and we excluded AC18 to make the plot more readable.}
\label{fig:distribution}
\end{figure}

\vspace{0.3cm}
In Figure \ref{fig:distribution}, we plot one representative dataset distribution per cluster  to show they are all Gaussian distributions but of different parameters.

\subsection{Prediction Scenarios}\label{sec:TrainingStrat}
We design and run our experiments in two main parts, composed of several prediction scenarios. To train the predictive models in each scenario, we follow a procedure similar to the one outlined in our previous work \cite{rossi2022bayesian} and explain its details in section \ref{sec:setup}. 

The first part of the experiments focuses on evaluating the uncertainty-aware predictive models, where we compare univariate and bivariate predictors trained and tested on one or twelve of the datasets described in the previous section for a total of four prediction scenarios. Scenarios are characterized by a specific prediction task (univariate or bivariate) and specific training data (single dataset or multiple datasets). Therefore, we distinguish each model by adding a specific prefix to its name: U and B for univariate and bivariate models, respectively, and S and M when trained on a single dataset or multiple (twelve) datasets, respectively. For example, M-B-HBNN refers to an HBNN trained with multiple datasets for a bivariate prediction. Also, it is worth noting that univariate models predict just one resource at a time, and bivariate models simultaneously predict both resources (i.e.\ processing units and memory). 
The goal of this experiment is twofold: understand if the joint prediction of the two targets outperforms the univariate case and quantify how impactful is the addition of different traces. Bivariate prediction is computationally harder, but it could give better results especially if the joint distribution can not be decomposed in 2 univariate ones.

\begin{figure*}[htbp]
\centerline{\includegraphics[width=\linewidth]{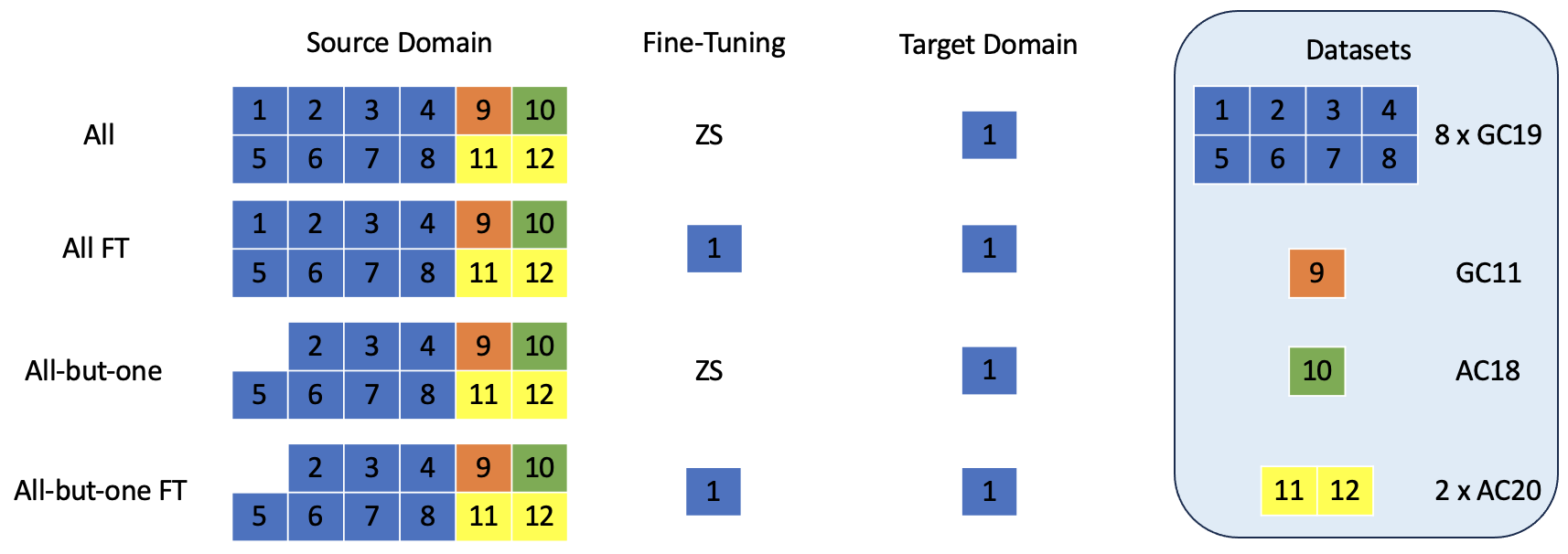}}
\caption{An illustrative example of four training scenarios for the second part of the experiments. In this example, we consider dataset number 1 of GC19 as the target domain (and FT dataset where applicable), and the source domains are reported accordingly. To note, ZS stands for Zero-Shot, so that, in practice, we do not fine-tune the model.}
\label{fig:prediction_scenarios}
\end{figure*}

The second part of the experiments investigates the domain generalization capabilities of the uncertainty-aware models. The methodology we describe can be applied to any of the models proposed in the first part of the experiments, but it is expensive. Thus, in this second part, we select only the M-B-HBNN network, being the best model accordingly to the results of the first part of the experiments (see section \ref{sec:results}). 

We design six prediction scenarios that differ based on the combinations of datasets used for training (i.e.\ source domain) and testing (i.e.\ target domain), see figure \ref{fig:prediction_scenarios} for an illustrative example of four of them. It is worth noting that we only build the following most representative scenarios among all the possible combinations we could build, and we leave exhaustive experimentation for future work.

\begin{itemize}
    \item \textbf{All}: the source domain is all the (twelve) datasets, and the target domain is one of the (twelve) datasets at a time. In other words, we use a portion of all the twelve source datasets (domains) for training, and we use another portion of one of these datasets (domains) for testing. The final performance is obtained by averaging results across all the twelve target datasets.
    \item \textbf{All FT}: we consider \textit{All} as the pretrained model. One by one, we use each of the (twelve) datasets to fine-tune \textit{All} and as a target domain. The final performance is obtained by averaging results across all the twelve target domains. This scenario assesses whether the FT process leads to a better weight configuration for the target domain we want to predict.
    \item \textbf{All-but-one}: the model is trained with all but one datasets (source domain). We then predict the remaining dataset (target domain). The final performance is obtained by averaging results across all the twelve target domains. This experiment further investigates the model's generalisation capabilities on unseen datasets in a zero-shot fashion. The real-world motivation of this scenario, for example, is being able to predict the workload of a new cluster from day one (instead of waiting a few days to collect historical workload data and train a model to predict this new cluster).
    \item \textbf{All-but-one FT}: we consider \textit{All-but-one} as the pretrained model. One by one, we use each of the excluded datasets to fine-tune \textit{All-but-one} and as a target domain. This investigates the combination of pretraining and FT on newly available data. The final performance is obtained by averaging results across all the twelve target domains.
    \item \textbf{TL-GC19}: we only use the eight datasets from GC19 and we build four sub-scenarios, i.e.\ \textit{GC19 All} resembling \textit{All}, \textit{GC19 All FT} resembling \textit{All FT}, \textit{GC19 All-but-one} resembling \textit{All-but-one}, \textit{GC19 All-but-one FT} resembling \textit{All-but-one FT}. These scenarios assess whether pretraining and FT enhance generalisation capabilities on different domains from the same providers.
    \item \textbf{TL-GC19vsOther}: the source domain is GC19 and we test its zero-shot and FT performance on GC11, AC18 and AC20. These scenarios assess whether pretraining and FT enhance generalisation capabilities on different domains from other providers or clusters from the same provider.

\end{itemize}

\subsection{Setup}\label{sec:setup}

Experiments are performed with a CPU Intel\textregistered Xeon\textregistered Gold 6240 at 2.60GHz and GPU NVIDIA Quadro RTX 8000 with 48 GB of memory where Ubuntu 20.04 is installed. For each dataset, the first 80\% data points are used as the training set, 20\% of which is used as the validation set, and the remaining 20\% as the test set. When training with multiple datasets, the training time frames of all datasets are aligned to ensure that one trace cannot exploit information of a particular timestamp in another trace. 

In the first part of the experiments, we tune the following hyperparameters with Talos library on the validation set: the number of neurons for each layer, the batch size, the activation functions, the learning rate, the number of kernels in the 1DConv layer and the optimizer with its momentum and decay coefficients. We train the models with their best hyperparameters ten times for each scenario using various random seeds as initialisation to assess the optimisation algorithm's convergence, and we use early stopping on the validation set to avoid overfitting. We forecast the 5-minute interval of demand 10 minutes in the future, where 10 minutes is sufficient for most real-world applications \cite{mao2010cloud, baldan2016forecasting}, e.g. resource allocation, vertical scaling etc. All three architectures are implemented in Keras; for the HBNN and LSTMD we also used TensorFlow probability. We share a GitHub repository that contains further information on the models' architecture and the search space for the hyperparameters. 

For the second part of the experiments, the datasets are split as in the first part, and we use the best hyperparameters found on the optimisation for the M-B-HBNN model. We present the results and the evaluation metrics in the next section.

\section{Results}\label{sec:results}

In this section, we evaluate the models in terms of point estimate accuracy, the efficiency of the predicted resources and runtime performance. In each subsection, we first assess the extension to bivariate models and those trained with multiple datasets. We then discuss the results based on the TL approach.

\subsection{On models comparison}


We first evaluate LSTM, LSTMD and HBNN regarding point estimate accuracy. The metrics employed for this evaluation are Mean Squared Error (MSE) and Mean Absolute Error (MAE), widely used to assess time series forecasting approaches. In the case of HBNN and LSTMD models, the error is computed w.r.t. the mean of the predicted distribution, while for the LSTM, the error is calculated based on the point prediction. Table \ref{tab:pointacc} shows the results of processing units and memory forecast for all models' combinations, i.e.\ single/multiple datasets and univariate/bivariate. Also, each model's performance is the average MSE and MAE computed across the twelve datasets.


First of all, results for MSE and MAE are consistent, so both metrics reveal the same patterns in the evaluation. Training a model with multiple datasets improves the accuracy of bivariate models, and the bigger improvement is for HBNN. The opposite happens for univariate models, even if performance degradation is quite limited. Also, univariate models outperform bivariate models, as confirmed by other previous works \cite{he2022multivariate}. This happens especially when the size of the overall training data is small, there is no strong correlation between time series \cite{du2003univariate} and when the focus is on short-term predictions \cite{chayama2016univariate}. Indeed, we analysed the Pearson correlation between the processing units and memory demand, finding a weak correlation between the time series and determining that homoscedasticity holds according to the Breusch-Pagan test \cite{breusch1979simple}. Considering the three architectures separately, LSTMD achieves the best accuracy score across the different training combinations. On the contrary, the S-B-HBNN fails to converge due to the extra complexity given by the Bayesian layer; the S-B-LSTMD struggles and does not achieve the same performance as the univariate, while S-B-LSTM worsens to a lesser extent. 

As the results stand, it is not clear whether training multivariate models (cheaper) with multiple datasets (more expensive) has a positive payoff in comparison with training more univariate models (more expensive) with a single dataset (cheaper). Moreover, accordingly to the Diebold-Mariano test \cite{diebold2002comparing} at 95\% confidence level\footnote{We run a pairwise test for each of the model's combinations}, the performances of all models are not statistically different (except for the single bivariate models, which are different from all other models). For these reasons, we cannot limit the analysis to these traditional accuracy metrics. 

\begin{table}[htbp]
\caption{Average MSE/MAE comparison for processing units and memory prediction. In bold, the best model overall, and in italics, the best model in their groups.}
\centering
\begin{tabular}{c|ll|ll|}
\cline{2-5}
 & \multicolumn{2}{c|}{\textbf{Processing units}} & \multicolumn{2}{c|}{\textbf{Memory}} \\ \hline
\multicolumn{1}{|c|}{\textbf{Model}} & \multicolumn{1}{c|}{\textbf{MSE}} & \multicolumn{1}{c|}{\textbf{MAE}} & \multicolumn{1}{c|}{\textbf{MSE}} & \multicolumn{1}{c|}{\textbf{MAE}} \\ \hline
\multicolumn{1}{|c|}{\textbf{S-U-LSTM}} & \multicolumn{1}{c|}{0.0041} & \multicolumn{1}{c|}{0.0457} & \multicolumn{1}{c|}{0.0042} & \multicolumn{1}{c|}{0.0425} \\ \hline
\multicolumn{1}{|c|}{\textbf{S-U-LSTMD}} & \multicolumn{1}{c|}{\textbf{0.0040}} & \multicolumn{1}{c|}{\textit{0.0455}} & \multicolumn{1}{c|}{\textbf{0.0041}} & \multicolumn{1}{c|}{\textbf{0.042}} \\ \hline
\multicolumn{1}{|c|}{\textbf{S-U-HBNN}} & \multicolumn{1}{c|}{0.0047} & \multicolumn{1}{c|}{0.0502} & \multicolumn{1}{c|}{0.0044} & \multicolumn{1}{c|}{0.0455} \\ \hline \hline
\multicolumn{1}{|c|}{\textbf{S-B-LSTM}} & \multicolumn{1}{l|}{\textit{0.0047}} & \textit{0.0500} & \multicolumn{1}{l|}{\textit{0.0054}} & \textit{0.0487} \\ \hline
\multicolumn{1}{|c|}{\textbf{S-B-LSTMD}} & \multicolumn{1}{l|}{0.0061} & 0.0556 & \multicolumn{1}{l|}{0.0088} & 0.0631 \\ \hline
\multicolumn{1}{|c|}{\textbf{S-B-HBNN}} & \multicolumn{1}{l|}{0.0299} & 0.1285 & \multicolumn{1}{l|}{0.0386} & 0.1533 \\ \hline \hline
\multicolumn{1}{|c|}{\textbf{M-U-LSTM}} & \multicolumn{1}{c|}{0.0044} & \multicolumn{1}{c|}{0.0474} & \multicolumn{1}{c|}{0.0048} & \multicolumn{1}{c|}{0.0447} \\ \hline
\multicolumn{1}{|c|}{\textbf{M-U-LSTMD}} & \multicolumn{1}{l|}{\textit{0.0041}} & \textit{0.0464} & \multicolumn{1}{l|}{\textit{0.0044}} & \textit{0.0439} \\ \hline
\multicolumn{1}{|c|}{\textbf{M-U-HBNN}} & \multicolumn{1}{l|}{0.0047} & 0.0485 & \multicolumn{1}{l|}{0.0052} & 0.0498 \\ \hline \hline
\multicolumn{1}{|c|}{\textbf{M-B-LSTM}} & \multicolumn{1}{l|}{0.0044} & 0.0479 & \multicolumn{1}{l|}{0.0044} & 0.0438 \\ \hline
\multicolumn{1}{|c|}{\textbf{M-B-LSTMD}} & \multicolumn{1}{l|}{0.0046} & \textbf{0.0446} & \multicolumn{1}{l|}{0.0046} & 0.0446 \\ \hline
\multicolumn{1}{|c|}{\textbf{M-B-HBNN}} & \multicolumn{1}{l|}{\textit{0.0042}} & 0.0471 & \multicolumn{1}{l|}{\textit{0.0043}} & \textit{0.0436} \\ \hline
\end{tabular}%
\label{tab:pointacc}
\end{table}

We further assess the models using the service level metrics defined in \cite{rossi2022bayesian}, i.e.\ where predictions are tailored to a specific target service level. In the case of HBNN and LSTMD, which predict a probability distribution, the final prediction values are computed w.r.t. the \textit{upper bound} (UB) of the confidence interval with a target service level varying from 90\% to 99.5\%. For LSTM, we could infer a confidence interval from its point estimate, i.e.\ the output of the network plus a fixed threshold, e.g. 5\%, as done in literature \cite{minarolli2014cross, minarolli2017tackling}. Instead, to avoid penalising the baseline too much, we calculate such an interval dynamically with the following procedure. We check the SR value scored by HBNN (LSTMD), and we adjust the interval to achieve the same SR value. In other words, we set the desired SR value first, and we calculate the interval accordingly.


The service level metrics that we measure are:
\begin{itemize}
    \item Success Rate (SR): the percentage of future demand within the confidence interval, i.e.\ whether the target confidence level is met. For example, when a 95\% service level is set, we would like 95\% of the requests to be within the UB of the prediction.
    
    \item Overprediction (OP): it is related to the predicted resource waste, i.e.\ the percentage amount of overprediction, defined as the difference between the UB of the prediction and the real demand for the requests within the confidence interval.
    
    \item Underprediction (UP): it quantifies the amount of unmatched demand that the customer requested, i.e.\ the percentage amount of underprediction, defined as the difference between the real demand and the UB of the prediction for the requests greater than the UB.
    
    \item Total Predicted Resources (TPR): it quantifies the overall amount of predicted resources that would be given as input to the resource manager, i.e.\ the sum of all the upper bounds of the predictions, in percentage w.r.t. the real demand.
\end{itemize}

\begin{table*}
\caption{Results for the service level metrics. We compare each HBNN and LSTMD with an LSTM baseline. Confidence intervals for LSTMs are tailored to the success rates achieved by HBNN and LSTMD. In bold we report the best model for each subgroup, and in italics the best model for each pair. }
\begin{center}
\begin{tabular}{cc|cccc||cccc|}
\cline{3-10}
 &  & \multicolumn{4}{c||}{\multirow{2}{*}{\textbf{Processing units}}} & \multicolumn{4}{c|}{\multirow{2}{*}{\textbf{Memory}}} \\
 &  & \multicolumn{4}{c||}{} & \multicolumn{4}{c|}{} \\ \hline
\multicolumn{1}{|c|}{\multirow{2}{*}{\textbf{Target QoS}}} & \multirow{2}{*}{\textbf{Model}} & \multicolumn{1}{c|}{\multirow{2}{*}{\textbf{SR (\%)}}} & \multicolumn{1}{c|}{\multirow{2}{*}{\textbf{OP (\%)}}} & \multicolumn{1}{c|}{\multirow{2}{*}{\textbf{UP (\%)}}} & \multirow{2}{*}{\textbf{TPR (\%)}} & \multicolumn{1}{c|}{\multirow{2}{*}{\textbf{SR (\%)}}} & \multicolumn{1}{c|}{\multirow{2}{*}{\textbf{OP (\%)}}} & \multicolumn{1}{c|}{\multirow{2}{*}{\textbf{UP (\%)}}} & \multirow{2}{*}{\textbf{TPR (\%)}} \\ 
\multicolumn{1}{|c|}{} &  & \multicolumn{1}{c|}{} & \multicolumn{1}{c|}{} & \multicolumn{1}{c|}{} &  & \multicolumn{1}{c|}{} & \multicolumn{1}{c|}{} & \multicolumn{1}{c|}{} &  \\ \hhline{==========}
\multicolumn{1}{|c|}{\multirow{16}{*}{95\%}} & S-U-LSTM & \multicolumn{1}{c|}{93.86} & \multicolumn{1}{c|}{\textit{18.08}} & \multicolumn{1}{c|}{0.53} & \textit{117.55} & \multicolumn{1}{c|}{93.54} & \multicolumn{1}{c|}{\textit{\textbf{12.58}}} & \multicolumn{1}{c|}{0.39} & \textit{\textbf{112.18}} \\ \cline{2-10}
\multicolumn{1}{|c|}{} & S-U-HBNN & \multicolumn{1}{c|}{93.86} & \multicolumn{1}{c|}{18.09} & \multicolumn{1}{c|}{\textit{0.51}} & 117.58 & \multicolumn{1}{c|}{93.54} & \multicolumn{1}{c|}{13.59} & \multicolumn{1}{c|}{\textit{0.38}} & 113.21 \\\hhline{|~=========} 
\multicolumn{1}{|c|}{} & S-U-LSTMD & \multicolumn{1}{c|}{93.14} & \multicolumn{1}{c|}{17.11} & \multicolumn{1}{c|}{\textit{0.56}} & 116.55 & \multicolumn{1}{c|}{93.26} & \multicolumn{1}{c|}{\textit{\textbf{11.80}}} & \multicolumn{1}{c|}{0.41} & \textit{\textbf{111.39}} \\ \cline{2-10} 
\multicolumn{1}{|c|}{} & S-U-LSTM & \multicolumn{1}{c|}{93.14} & \multicolumn{1}{c|}{\textit{\textbf{16.57}}} & \multicolumn{1}{c|}{0.63} & \textit{\textbf{115.94}} & \multicolumn{1}{c|}{93.26} & \multicolumn{1}{c|}{12.43} & \multicolumn{1}{c|}{\textit{0.40}} & 112.03 \\ \hhline{|~=========} \hhline{|~=========} 
\multicolumn{1}{|c|}{} & S-B-LSTM & \multicolumn{1}{c|}{91.93} & \multicolumn{1}{c|}{85.88} & \multicolumn{1}{c|}{3.27} & 182.61 & \multicolumn{1}{c|}{91.89} & \multicolumn{1}{c|}{71.49} & \multicolumn{1}{c|}{\textit{1.33}} & 170.16 \\ \cline{2-10} 
\multicolumn{1}{|c|}{} & S-B-HBNN & \multicolumn{1}{c|}{91.93} & \multicolumn{1}{c|}{\textit{82.10}} & \multicolumn{1}{c|}{\textit{2.93}} & \textit{179.17} & \multicolumn{1}{c|}{91.89} & \multicolumn{1}{c|}{\textit{71.47}} & \multicolumn{1}{c|}{1.43} & \textit{170.04} \\ \hhline{|~=========} 
\multicolumn{1}{|c|}{} & S-B-LSTMD & \multicolumn{1}{c|}{92.61} & \multicolumn{1}{c|}{\textit{84.26}} & \multicolumn{1}{c|}{3.01} & \textit{181.25} & \multicolumn{1}{c|}{92.45} & \multicolumn{1}{c|}{\textit{71.89}} & \multicolumn{1}{c|}{1.39} & \textit{170.50} \\ \cline{2-10} 
\multicolumn{1}{|c|}{} & S-B-LSTM & \multicolumn{1}{c|}{92.61} & \multicolumn{1}{c|}{89.09} & \multicolumn{1}{c|}{\textit{3.00}} & 186.08 & \multicolumn{1}{c|}{92.45} & \multicolumn{1}{c|}{72.58} & \multicolumn{1}{c|}{\textit{1.24}} & 171.34 \\ \hhline{|~=========} \hhline{|~=========} 
\multicolumn{1}{|c|}{} & M-U-LSTM & \multicolumn{1}{c|}{91.41} & \multicolumn{1}{c|}{15.63} & \multicolumn{1}{c|}{0.76} & 114.87 & \multicolumn{1}{c|}{95.78} & \multicolumn{1}{c|}{15.54} & \multicolumn{1}{c|}{\textit{0.27}} & 115.27 \\ \cline{2-10} 
\multicolumn{1}{|c|}{} & M-U-HBNN & \multicolumn{1}{c|}{91.41} & \multicolumn{1}{c|}{\textbf{\textit{15.39}}} & \multicolumn{1}{c|}{\textit{0.70}} & \textit{\textbf{114.69}} & \multicolumn{1}{c|}{95.78} & \multicolumn{1}{c|}{\textit{15.20}} & \multicolumn{1}{c|}{0.30} & \textit{114.90} \\ \hhline{|~=========} 
\multicolumn{1}{|c|}{} & M-U-LSTMD & \multicolumn{1}{c|}{94.62} & \multicolumn{1}{c|}{\textit{19.55}} & \multicolumn{1}{c|}{\textit{0.43}} & \textit{119.12} & \multicolumn{1}{c|}{95.18} & \multicolumn{1}{c|}{\textit{14.53}} & \multicolumn{1}{c|}{0.32} & \textit{114.21} \\ \cline{2-10} 
\multicolumn{1}{|c|}{} & M-U-LSTM & \multicolumn{1}{c|}{94.62} & \multicolumn{1}{c|}{19.67} & \multicolumn{1}{c|}{0.46} & 119.21 & \multicolumn{1}{c|}{95.18} & \multicolumn{1}{c|}{14.91} & \multicolumn{1}{c|}{\textit{0.30}} & 114.61 \\ \hhline{|~=========} \hhline{|~=========} 
\multicolumn{1}{|c|}{} & M-B-LSTM & \multicolumn{1}{c|}{95.70} & \multicolumn{1}{c|}{22.61} & \multicolumn{1}{c|}{0.39} & 122.22 & \multicolumn{1}{c|}{97.39} & \multicolumn{1}{c|}{19.90} & \multicolumn{1}{c|}{\textit{\textbf{0.14}}} & 119.76 \\ \cline{2-10} 
\multicolumn{1}{|c|}{} & M-B-HBNN & \multicolumn{1}{c|}{95.70} & \multicolumn{1}{c|}{\textit{21.58}} & \multicolumn{1}{c|}{\textit{\textbf{0.38}}} & \textit{121.20} & \multicolumn{1}{c|}{97.39} & \multicolumn{1}{c|}{\textit{19.47}} & \multicolumn{1}{c|}{0.16} & \textit{119.31} \\ \hhline{|~=========} 
\multicolumn{1}{|c|}{} & M-B-LSTMD & \multicolumn{1}{c|}{96.99} & \multicolumn{1}{c|}{28.25} & \multicolumn{1}{c|}{\textit{\textbf{0.24}}} & 128.01 & \multicolumn{1}{c|}{99.00} & \multicolumn{1}{c|}{\textit{24.30}} & \multicolumn{1}{c|}{\textit{\textbf{0.06}}} & 124.24 \\ \cline{2-10} 
\multicolumn{1}{|c|}{} & M-B-LSTM & \multicolumn{1}{c|}{96.99} & \multicolumn{1}{c|}{\textit{25.84}} & \multicolumn{1}{c|}{0.27} & \textit{125.57 }& \multicolumn{1}{c|}{99.00} & \multicolumn{1}{c|}{\textit{24.30}} & \multicolumn{1}{c|}{0.06} & \textit{124.24} \\ \hline \hline \hline
\multicolumn{1}{|c|}{\multirow{16}{*}{97\%}} & S-U-LSTM & \multicolumn{1}{c|}{95.22} & \multicolumn{1}{c|}{\textit{20.07}} & \multicolumn{1}{c|}{0.41} & \textit{119.66} & \multicolumn{1}{c|}{95.06} & \multicolumn{1}{c|}{\textit{\textbf{13.92}}} & \multicolumn{1}{c|}{0.31} & \textit{\textbf{113.61}} \\ \cline{2-10}
\multicolumn{1}{|c|}{} & S-U-HBNN & \multicolumn{1}{c|}{95.22} & \multicolumn{1}{c|}{20.13} & \multicolumn{1}{c|}{\textit{0.39}} & 119.79 & \multicolumn{1}{c|}{95.06} & \multicolumn{1}{c|}{15.22} & \multicolumn{1}{c|}{\textit{0.28}} & 114.94 \\ \hhline{|~=========} 
\multicolumn{1}{|c|}{} & S-U-LSTMD & \multicolumn{1}{c|}{94.58} & \multicolumn{1}{c|}{\textit{\textbf{19.13}}} & \multicolumn{1}{c|}{\textit{0.43}} & \textit{\textbf{118.69}} & \multicolumn{1}{c|}{94.86} & \multicolumn{1}{c|}{\textit{\textbf{13.31}}} & \multicolumn{1}{c|}{\textit{0.31}} & \textit{\textbf{113.00}} \\ \cline{2-10} 
\multicolumn{1}{|c|}{} & S-U-LSTM & \multicolumn{1}{c|}{94.58} & \multicolumn{1}{c|}{19.36} & \multicolumn{1}{c|}{0.45} & 118.91 & \multicolumn{1}{c|}{94.86} & \multicolumn{1}{c|}{13.79} & \multicolumn{1}{c|}{0.32} & 113.47 \\ \hhline{|~=========} \hhline{|~=========} 

\multicolumn{1}{|c|}{} & S-B-LSTM & \multicolumn{1}{c|}{93.46} & \multicolumn{1}{c|}{95.31} & \multicolumn{1}{c|}{2.54} & 192.78 & \multicolumn{1}{c|}{94.78} & \multicolumn{1}{c|}{\textit{79.89}} & \multicolumn{1}{c|}{0.77} & \textit{179.12} \\ \cline{2-10} 
\multicolumn{1}{|c|}{} & S-B-HBNN & \multicolumn{1}{c|}{93.46} & \multicolumn{1}{c|}{\textit{91.38}} & \multicolumn{1}{c|}{\textit{2.21}} & \textit{189.16} & \multicolumn{1}{c|}{94.78} & \multicolumn{1}{c|}{81.18} & \multicolumn{1}{c|}{\textit{0.75}} & 180.43 \\ \hhline{|~=========} 
\multicolumn{1}{|c|}{} & S-B-LSTMD & \multicolumn{1}{c|}{93.58} & \multicolumn{1}{c|}{\textit{93.57}} & \multicolumn{1}{c|}{\textit{2.34}} & \textit{191.27} & \multicolumn{1}{c|}{94.58} & \multicolumn{1}{c|}{80.98} & \multicolumn{1}{c|}{\textit{0.78}} & 180.20 \\ \cline{2-10} 
\multicolumn{1}{|c|}{} & S-B-LSTM & \multicolumn{1}{c|}{93.58} & \multicolumn{1}{c|}{94.22} & \multicolumn{1}{c|}{2.49} & 193.52 & \multicolumn{1}{c|}{94.58} & \multicolumn{1}{c|}{\textit{77.64}} & \multicolumn{1}{c|}{0.90} & \textit{176.74} \\ \hhline{|~=========} \hhline{|~=========} 
\multicolumn{1}{|c|}{} & M-U-LSTM & \multicolumn{1}{c|}{93.82} & \multicolumn{1}{c|}{18.27} & \multicolumn{1}{c|}{0.55} & 117.72 & \multicolumn{1}{c|}{96.91} & \multicolumn{1}{c|}{17.44} & \multicolumn{1}{c|}{\textit{0.20}} & 117.24 \\ \cline{2-10} 
\multicolumn{1}{|c|}{} & M-U-HBNN & \multicolumn{1}{c|}{93.82} & \multicolumn{1}{c|}{\textit{\textbf{17.91}}} & \multicolumn{1}{c|}{\textit{0.50}} & \textit{\textbf{117.41}} & \multicolumn{1}{c|}{96.91} & \multicolumn{1}{c|}{\textit{17.14}} & \multicolumn{1}{c|}{0.23} & \textit{116.91} \\ \hhline{|~=========}  
\multicolumn{1}{|c|}{} & M-U-LSTMD & \multicolumn{1}{c|}{95.95} & \multicolumn{1}{c|}{21.80} & \multicolumn{1}{c|}{\textit{0.31}} & 121.49 & \multicolumn{1}{c|}{96.39} & \multicolumn{1}{c|}{\textit{16.36}} & \multicolumn{1}{c|}{0.25} & \textit{116.11} \\ \cline{2-10} 
\multicolumn{1}{|c|}{} & M-U-LSTM & \multicolumn{1}{c|}{95.95} & \multicolumn{1}{c|}{\textit{21.56}} & \multicolumn{1}{c|}{0.37} & \textit{121.19} & \multicolumn{1}{c|}{96.39} & \multicolumn{1}{c|}{16.68} & \multicolumn{1}{c|}{\textit{0.23}} & 116.45 \\ \hhline{|~=========} \hhline{|~=========} 
\multicolumn{1}{|c|}{} & M-B-LSTM & \multicolumn{1}{c|}{96.75} & \multicolumn{1}{c|}{25.36} & \multicolumn{1}{c|}{0.28} & 125.07 & \multicolumn{1}{c|}{98.03} & \multicolumn{1}{c|}{\textit{21.18}} & \multicolumn{1}{c|}{\textit{\textbf{0.11}}} & \textit{121.07} \\ \cline{2-10} 
\multicolumn{1}{|c|}{} & M-B-HBNN & \multicolumn{1}{c|}{96.75} & \multicolumn{1}{c|}{\textit{24.18}} & \multicolumn{1}{c|}{\textit{\textbf{0.28}}} & \textit{123.90} & \multicolumn{1}{c|}{98.03} & \multicolumn{1}{c|}{21.80} & \multicolumn{1}{c|}{0.11} & 121.69 \\ \hhline{|~=========} 
\multicolumn{1}{|c|}{} & M-B-LSTMD & \multicolumn{1}{c|}{97.75} & \multicolumn{1}{c|}{31.51} & \multicolumn{1}{c|}{\textit{\textbf{0.16}}} & 131.35 & \multicolumn{1}{c|}{99.44} & \multicolumn{1}{c|}{\textit{27.38}} & \multicolumn{1}{c|}{0.03} & \textit{127.35} \\ \cline{2-10} 
\multicolumn{1}{|c|}{} & M-B-LSTM & \multicolumn{1}{c|}{97.75} & \multicolumn{1}{c|}{\textit{29.22}} & \multicolumn{1}{c|}{0.18} & \textit{129.04} & \multicolumn{1}{c|}{99.44} & \multicolumn{1}{c|}{29.26} & \multicolumn{1}{c|}{\textit{\textbf{0.02}}} & 129.24 \\ \hline \hline \hline\multicolumn{1}{|c|}{\multirow{16}{*}{99\%}} & S-U-LSTM & \multicolumn{1}{c|}{97.03} & \multicolumn{1}{c|}{24.61} & \multicolumn{1}{c|}{0.23} & 124.37 & \multicolumn{1}{c|}{97.55} & \multicolumn{1}{c|}{\textit{\textbf{18.11}}} & \multicolumn{1}{c|}{\textit{0.16}} & \textit{\textbf{117.95}} \\ \cline{2-10}
\multicolumn{1}{|c|}{} & S-U-HBNN & \multicolumn{1}{c|}{97.03} & \multicolumn{1}{c|}{\textit{24.04}} & \multicolumn{1}{c|}{\textit{0.23}} & \textit{123.81} & \multicolumn{1}{c|}{97.55} & \multicolumn{1}{c|}{18.37} & \multicolumn{1}{c|}{0.16} & 118.21 \\ \hhline{|~=========} 
\multicolumn{1}{|c|}{} & S-U-LSTMD & \multicolumn{1}{c|}{96.35} & \multicolumn{1}{c|}{23.01} & \multicolumn{1}{c|}{\textit{0.27}} & 122.75 & \multicolumn{1}{c|}{97.07} & \multicolumn{1}{c|}{\textit{\textbf{16.22}}} & \multicolumn{1}{c|}{\textit{0.19}} & \textit{\textbf{116.03}} \\ \cline{2-10} 
\multicolumn{1}{|c|}{} & S-U-LSTM & \multicolumn{1}{c|}{96.35} & \multicolumn{1}{c|}{\textit{\textbf{22.80}}} & \multicolumn{1}{c|}{0.29} & \textit{\textbf{122.51}} & \multicolumn{1}{c|}{97.07} & \multicolumn{1}{c|}{16.83} & \multicolumn{1}{c|}{0.19} & 116.64 \\ \hhline{|~=========} \hhline{|~=========} 

\multicolumn{1}{|c|}{} & S-B-LSTM & \multicolumn{1}{c|}{94.58} & \multicolumn{1}{c|}{113.27} & \multicolumn{1}{c|}{1.40} & 211.86 & \multicolumn{1}{c|}{98.07} & \multicolumn{1}{c|}{\textit{95.62}} & \multicolumn{1}{c|}{0.12} & \textit{195.50} \\ \cline{2-10} 
\multicolumn{1}{|c|}{} & S-B-HBNN & \multicolumn{1}{c|}{94.58} & \multicolumn{1}{c|}{\textit{109.11}} & \multicolumn{1}{c|}{\textit{1.08}} & \textit{208.03} & \multicolumn{1}{c|}{98.07} & \multicolumn{1}{c|}{100.15} & \multicolumn{1}{c|}{\textit{0.15}} & 200.00 \\ \hhline{|~=========}  
\multicolumn{1}{|c|}{} & S-B-LSTMD & \multicolumn{1}{c|}{94.42} & \multicolumn{1}{c|}{111.33} & \multicolumn{1}{c|}{\textit{1.23}} & 210.10 & \multicolumn{1}{c|}{97.91} & \multicolumn{1}{c|}{98.71} & \multicolumn{1}{c|}{\textit{0.16}} & 198.55 \\ \cline{2-10} 
\multicolumn{1}{|c|}{} & S-B-LSTM & \multicolumn{1}{c|}{94.42} & \multicolumn{1}{c|}{\textit{111.16}} & \multicolumn{1}{c|}{1.53} & \textit{209.63} & \multicolumn{1}{c|}{97.91} & \multicolumn{1}{c|}{\textit{94.98}} & \multicolumn{1}{c|}{0.15} & \textit{194.83} \\ \hhline{|~=========} \hhline{|~=========} 
\multicolumn{1}{|c|}{} & M-U-LSTM & \multicolumn{1}{c|}{96.47} & \multicolumn{1}{c|}{\textit{\textbf{22.75}}} & \multicolumn{1}{c|}{0.32} & \textit{\textbf{122.43}} & \multicolumn{1}{c|}{98.07} & \multicolumn{1}{c|}{\textit{20.65}} & \multicolumn{1}{c|}{\textit{0.16}} & \textit{120.49} \\ \cline{2-10} 
\multicolumn{1}{|c|}{} & M-U-HBNN & \multicolumn{1}{c|}{96.47} & \multicolumn{1}{c|}{22.80} & \multicolumn{1}{c|}{\textit{0.27}} & 122.53 & \multicolumn{1}{c|}{98.07} & \multicolumn{1}{c|}{20.85} & \multicolumn{1}{c|}{0.16} & 120.69 \\ \hhline{|~=========}  
\multicolumn{1}{|c|}{} & M-U-LSTMD & \multicolumn{1}{c|}{97.79} & \multicolumn{1}{c|}{\textit{26.14}} & \multicolumn{1}{c|}{\textit{0.17}} & \textit{125.96} & \multicolumn{1}{c|}{97.63} & \multicolumn{1}{c|}{19.86} & \multicolumn{1}{c|}{0.19} & 119.67 \\ \cline{2-10} 
\multicolumn{1}{|c|}{} & M-U-LSTM & \multicolumn{1}{c|}{97.79} & \multicolumn{1}{c|}{27.44} & \multicolumn{1}{c|}{0.18} & 127.26 & \multicolumn{1}{c|}{97.63} & \multicolumn{1}{c|}{\textit{19.37}} & \multicolumn{1}{c|}{\textit{0.19}} & \textit{119.18} \\ \hhline{|~=========} \hhline{|~=========} 
\multicolumn{1}{|c|}{} & M-B-LSTM & \multicolumn{1}{c|}{97.95} & \multicolumn{1}{c|}{29.95} & \multicolumn{1}{c|}{0.16} & 129.78 & \multicolumn{1}{c|}{99.04} & \multicolumn{1}{c|}{\textit{25.86}} & \multicolumn{1}{c|}{\textit{\textbf{0.04}}} & \textit{125.82} \\ \cline{2-10} 
\multicolumn{1}{|c|}{} & M-B-HBNN & \multicolumn{1}{c|}{97.95} & \multicolumn{1}{c|}{\textit{29.15}} & \multicolumn{1}{c|}{\textit{\textbf{0.15}}} & \textit{128.99} & \multicolumn{1}{c|}{99.04} & \multicolumn{1}{c|}{26.23} & \multicolumn{1}{c|}{0.06} & 126.17 \\ \hhline{|~=========} 
\multicolumn{1}{|c|}{} & M-B-LSTMD & \multicolumn{1}{c|}{99.04} & \multicolumn{1}{c|}{37.73} & \multicolumn{1}{c|}{\textit{\textbf{0.06}}} & 137.67 & \multicolumn{1}{c|}{99.80} & \multicolumn{1}{c|}{\textit{33.22}} & \multicolumn{1}{c|}{0.01} & \textit{133.21} \\ \cline{2-10} 
\multicolumn{1}{|c|}{} & M-B-LSTM & \multicolumn{1}{c|}{99.04} & \multicolumn{1}{c|}{\textit{36.54}} & \multicolumn{1}{c|}{0.06} & \textit{136.48} & \multicolumn{1}{c|}{99.80} & \multicolumn{1}{c|}{33.99} & \multicolumn{1}{c|}{\textit{\textbf{0.01}}} & 133.98 \\ \hline 
\end{tabular}
\label{tab:statistics}
\end{center}
\end{table*}

\begin{figure}[htbp]
\centerline{\includegraphics[width=\linewidth]{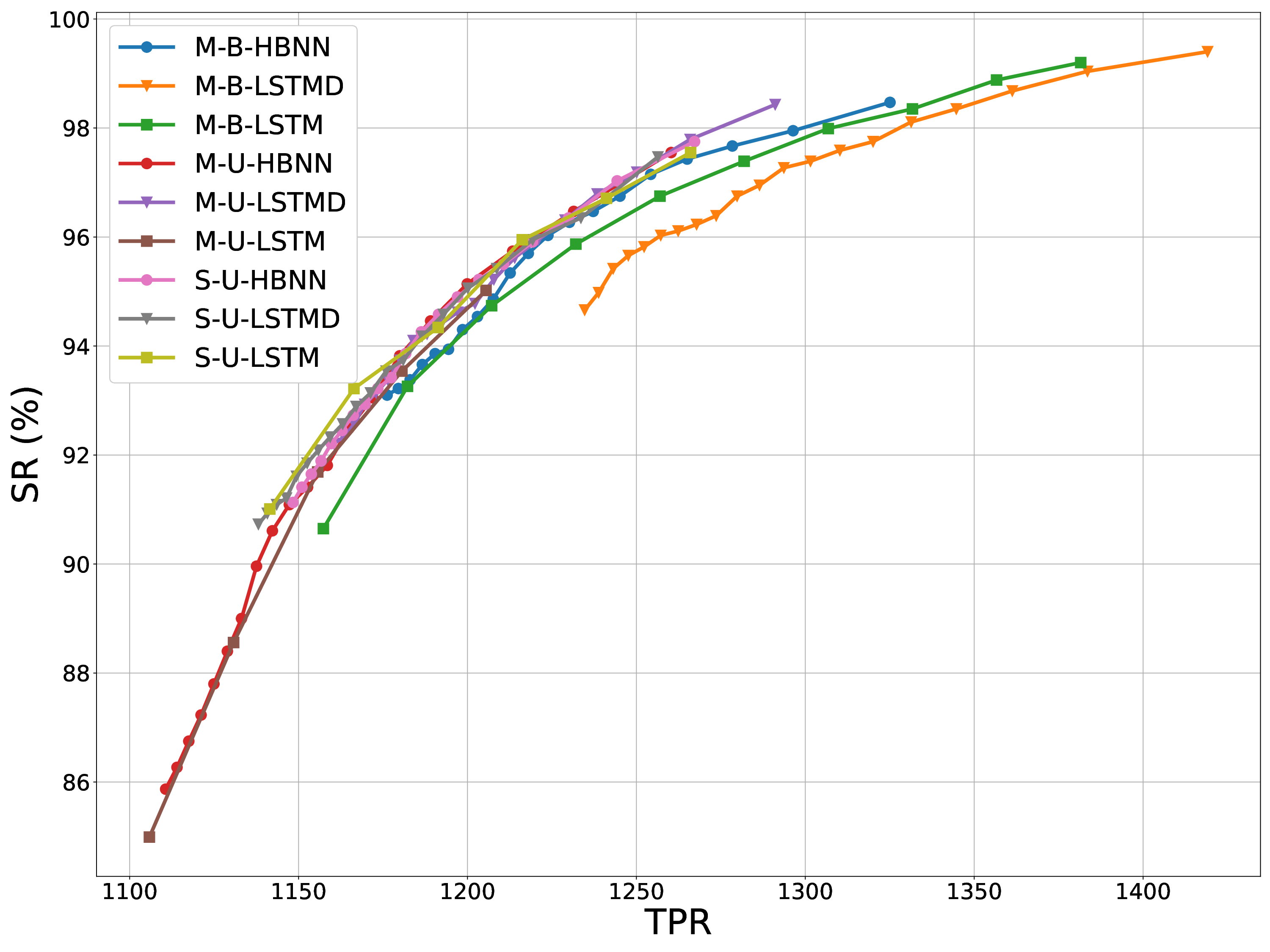}}
\caption{Total predicted processing units for S-U, M-U and M-B models. The closer the curve to the top left corner, the better the model.}
\label{fig:tprcpu}
\end{figure}

\begin{figure}[htbp]
\centerline{\includegraphics[width=\linewidth]{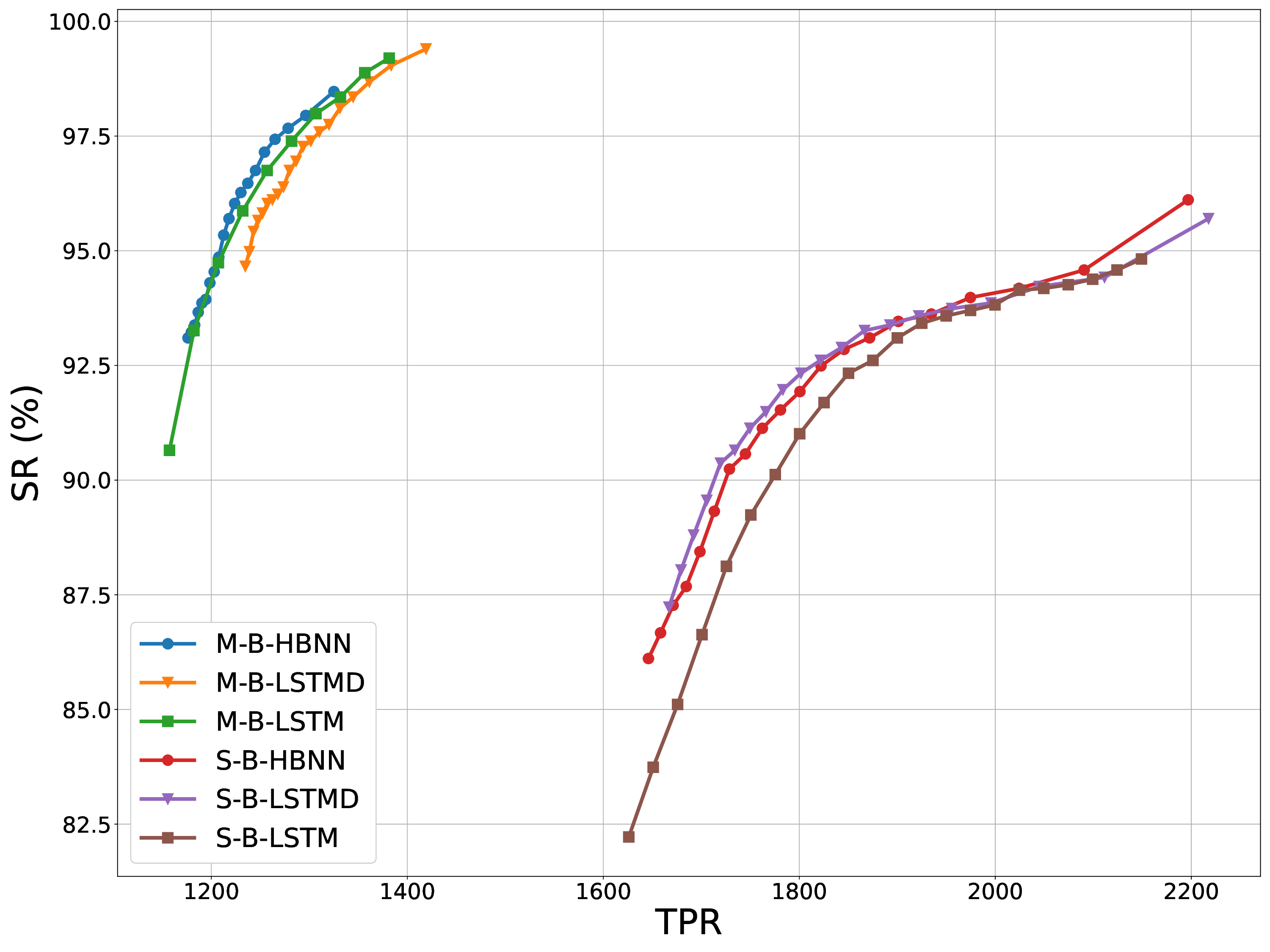}}
\caption{Total predicted processing units for bivariate models. The ones trained with a single dataset perform very poorly, with a much higher amount of TPR w.r.t. than the models trained with multiple datasets.}
\label{fig:tprcpubi}
\end{figure}

Table \ref{tab:statistics} shows values for the service level metrics, where we report only target service levels such as 95\%, 97\% and 99\% (as more significant for systems that require high-quality service levels). To help readers interpret such results, we also draw a graphical representation of the TPR versus the SR for processing units demand in Fig. \ref{fig:tprcpu}. The graph is drawn by computing the SR and TPR for the entire range of target service levels. One model outperforms another one when its curve is more to the left than the other's (SRs being equal) and its curve is above the other's (TPRs being equal). To improve the readability of the graph, we removed the curves of the single bivariate models, which achieve poor performance under this metric. From the aforementioned table and figure, we can see that the uncertainty-aware models consistently outperform the LSTM baseline, with HBNN outperforming LSTMD for the processing units and the other way around for the memory. Also, training with multiple datasets has more advantages over training with single datasets, despite the single univariate model seeming superior in the case of memory prediction at a 99\% service level. This confirms that the univariate prediction is easier than the bivariate version. Fig. \ref{fig:tprcpubi} focuses on the comparison of bivariate models trained with single and multiple datasets and confirms the results of the accuracy metrics, i.e.\ that the model benefits significantly if it is trained with more data. In particular, the M-B models achieve between around 25\% and 60\% saving in terms of TPR for both processing units and memory demand, with a higher SR.

\begin{table}[]
\centering
\caption{Average MSE and MAE comparison for resource prediction accuracy. The MSE and MAE in computed between the SR achieved by the models and the confidence level. In bold, the model overall.}
\begin{tabular}{c|cc|cc|}
\cline{2-5}
 & \multicolumn{2}{c|}{\textbf{Processing units}} & \multicolumn{2}{c|}{\textbf{Memory}} \\ \hline
\multicolumn{1}{|c|}{\textbf{Model}} & \multicolumn{1}{c|}{\textbf{MSE}} & \multicolumn{1}{c|}{\textbf{MAE}} & \multicolumn{1}{c|}{\textbf{MSE}} & \multicolumn{1}{c|}{\textbf{MAE}} \\ \hline
\multicolumn{1}{|c|}{\textbf{S-U-LSTMD}} & \multicolumn{1}{c|}{2.96} & \multicolumn{1}{c|}{1.47} & \multicolumn{1}{c|}{3.07} & \multicolumn{1}{c|}{1.67} \\ \hline
\multicolumn{1}{|c|}{\textbf{S-U-HBNN}} & \multicolumn{1}{c|}{1.69} & \multicolumn{1}{c|}{1.10} & \multicolumn{1}{c|}{1.80} & \multicolumn{1}{c|}{1.27} \\ \hline\hline
\multicolumn{1}{|c|}{\textbf{S-B-LSTMD}} & \multicolumn{1}{c|}{8.15} & 2.66 & \multicolumn{1}{c|}{6.51} & 2.34 \\ \hline
\multicolumn{1}{|c|}{\textbf{S-B-HBNN}} & \multicolumn{1}{c|}{11.80} & 3.32 & \multicolumn{1}{c|}{11.81} & 3.03 \\ \hline\hline
\multicolumn{1}{|c|}{\textbf{M-U-LSTMD}} & \multicolumn{1}{c|}{\textbf{1.24}} & \textbf{0.98} & \multicolumn{1}{c|}{\textbf{1.14}} & \textbf{0.92} \\ \hline
\multicolumn{1}{|c|}{\textbf{M-U-HBNN}} & \multicolumn{1}{c|}{12.07} & 3.34 & \multicolumn{1}{c|}{2.42} & 1.25 \\ \hline\hline
\multicolumn{1}{|c|}{\textbf{M-B-LSTMD}} & \multicolumn{1}{c|}{6.89} & 2.08 & \multicolumn{1}{c|}{21.05} & 3.94 \\ \hline
\multicolumn{1}{|c|}{\textbf{M-B-HBNN}} & \multicolumn{1}{c|}{1.97} & 1.11 & \multicolumn{1}{c|}{8.66} & 2.39 \\ \hline
\end{tabular}%
\label{tab:accuracyerrors}
\end{table}

\begin{figure}[htbp]
\centerline{\includegraphics[width=\linewidth]{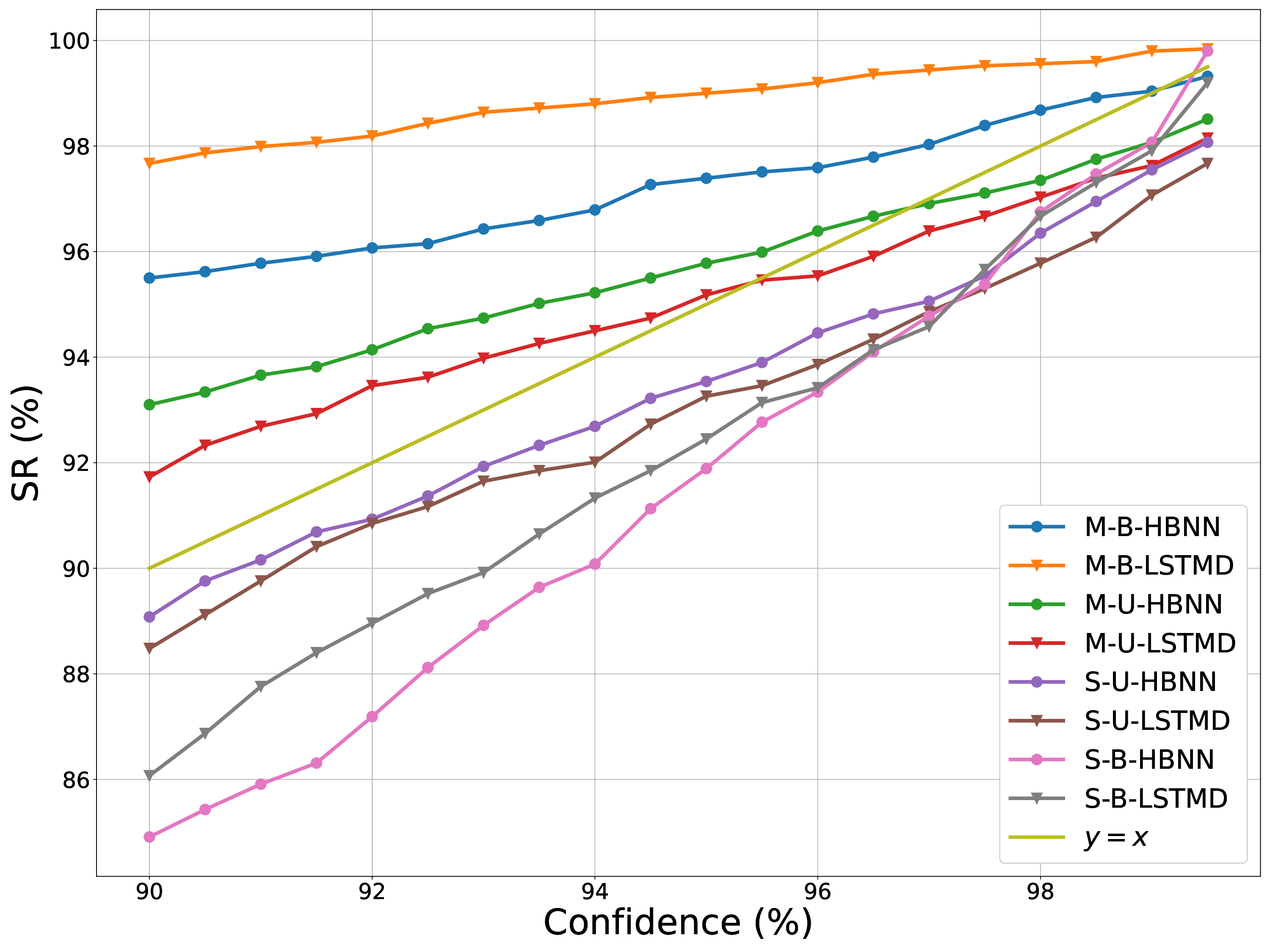}}
\caption{Target service level versus SR for memory prediction. The plot is built w.r.t. the UB of the confidence interval by varying the target service level. The closer the plot is to the line $y=x$, the more accurate the model under this metric.}
\label{fig:accmem}
\end{figure}

Moreover, we evaluate the impact of uncertainty-aware predictions by plotting the confidence levels versus the success rate achieved by the models. A perfect model would achieve an SR equal to the targeted confidence level, i.e.\ would resemble the bisector line $y=x$. Indeed, we also compute the MSE and MAE of each model's curve w.r.t. to such a bisector line to represent the plot in numerical values and enhance interpretation. The analysis is depicted in Fig. \ref{fig:accmem} and Table \ref{tab:accuracyerrors}. It shows M-U-LSTMD is the model that overall achieves the best accuracy, but there are also interesting differences between each combination of the model, e.g.\ the HBNN outperforms the LSTMD counterpart in the case of S-U and M-B versions. We should also consider that MSE and MAE are symmetric metrics. Depending on the target service level provided to the final customer, the cloud manager could prefer models that achieve a higher SR than the target confidence level rather than a lower one, even if the MSE and MAE are lower. For instance, we should favour a model that achieves a 96\% SR compared to 94\% w.r.t. a 95\% confidence level, despite the score being the same in terms of MSE and MAE. For this reason, asymmetric metrics would be more suitable to evaluate the model's accuracy combined with the confidence of the prediction.

\begin{table*}[htbp]
\begin{center}
\caption{Runtime performance analysis. In italics is the best model for each subgroup. The training time is computed by training the model with 20, 40, 60 and 80\%, respectively. The fine-tuning time is computed by fine-tuning the network every 6, 12, 18 and 24 steps, which corresponds to 30, 60, 90 and 120 minutes, respectively. The inference time is the time to predict one sample.}
\begin{tabular}{c|cccc|cccc|c|}
\cline{2-10}
 & \multicolumn{4}{c|}{ \makecell{\textbf{Training} \\ \textbf{time [s]}}} & \multicolumn{4}{c|}{ \makecell{\textbf{Fine-Tuning} \\ \textbf{time [s]}}} &  \makecell{\textbf{Inference} \\ \textbf{time [s]}} \\ \hline
\multicolumn{1}{|c|}{\textbf{Model}} & \multicolumn{1}{c|}{20\%} & \multicolumn{1}{c|}{40\%} & \multicolumn{1}{c|}{60\%} & 80\% & \multicolumn{1}{c|}{6} & \multicolumn{1}{c|}{12} & \multicolumn{1}{c|}{18} & 24 & 1 sample \\ \hline
\multicolumn{1}{|c|}{\textbf{S-U-LSTM}} & \multicolumn{1}{c|}{\textit{12}}& \multicolumn{1}{c|}{\textit{21}}& \multicolumn{1}{c|}{\textit{29}}& \multicolumn{1}{c|}{\textit{48}}& \multicolumn{1}{c|}{5.38}& \multicolumn{1}{c|}{7.4}& \multicolumn{1}{c|}{12.77}& \multicolumn{1}{c|}{9.29}& \multicolumn{1}{c|}{0.0002} \\ \hline
\multicolumn{1}{|c|}{\textbf{S-U-HBNN}} & \multicolumn{1}{c|}{35}& \multicolumn{1}{c|}{140}& \multicolumn{1}{c|}{268}& \multicolumn{1}{c|}{325}& \multicolumn{1}{c|}{\textit{1.57}}& \multicolumn{1}{c|}{\textit{1.63}}& \multicolumn{1}{c|}{\textit{1.7}}& \multicolumn{1}{c|}{\textit{1.64}}& \multicolumn{1}{c|}{0.0058} \\ \hline
\multicolumn{1}{|c|}{\textbf{S-U-LSTMD}} & \multicolumn{1}{c|}{15}& \multicolumn{1}{c|}{25}& \multicolumn{1}{c|}{38}& \multicolumn{1}{c|}{52}& \multicolumn{1}{c|}{2.49}& \multicolumn{1}{c|}{2.32}& \multicolumn{1}{c|}{2.22}& \multicolumn{1}{c|}{1.68}& \multicolumn{1}{c|}{\textit{0.00002}} \\ \hline \hline
\multicolumn{1}{|c|}{\textbf{S-B-LSTM}} & \multicolumn{1}{c|}{11}& \multicolumn{1}{c|}{21}& \multicolumn{1}{c|}{32}& \multicolumn{1}{c|}{\textit{53}}& \multicolumn{1}{c|}{3.03}& \multicolumn{1}{c|}{\textit{1.61}}& \multicolumn{1}{c|}{\textit{1.34}}& \multicolumn{1}{c|}{\textit{1.54}}& \multicolumn{1}{c|}{0.0002} \\ \hline
\multicolumn{1}{|c|}{\textbf{S-B-HBNN}} & \multicolumn{1}{c|}{57}& \multicolumn{1}{c|}{106}& \multicolumn{1}{c|}{154}& \multicolumn{1}{c|}{172}& \multicolumn{1}{c|}{\textit{1.84}}& \multicolumn{1}{c|}{2.01}& \multicolumn{1}{c|}{1.88}& \multicolumn{1}{c|}{2.01}& \multicolumn{1}{c|}{\textit{0.0001}} \\ \hline
\multicolumn{1}{|c|}{\textbf{S-B-LSTMD}} & \multicolumn{1}{c|}{\textit{10}}& \multicolumn{1}{c|}{\textit{16}}& \multicolumn{1}{c|}{\textit{25}}& \multicolumn{1}{c|}{237}& \multicolumn{1}{c|}{3.17}& \multicolumn{1}{c|}{4.06}& \multicolumn{1}{c|}{2.98}& \multicolumn{1}{c|}{2.26}& \multicolumn{1}{c|}{\textit{0.0001}} \\ \hline \hline
\multicolumn{1}{|c|}{\textbf{M-U-LSTM}} & \multicolumn{1}{c|}{1121}& \multicolumn{1}{c|}{947}& \multicolumn{1}{c|}{869}& \multicolumn{1}{c|}{1050}& \multicolumn{1}{c|}{2.54}& \multicolumn{1}{c|}{2.44}& \multicolumn{1}{c|}{2.37}& \multicolumn{1}{c|}{3.41}& \multicolumn{1}{c|}{\textit{0.0001}} \\ \hline
\multicolumn{1}{|c|}{\textbf{M-U-HBNN}} & \multicolumn{1}{c|}{\textit{486}}& \multicolumn{1}{c|}{\textit{425}}& \multicolumn{1}{c|}{\textit{530}}& \multicolumn{1}{c|}{\textit{621}}& \multicolumn{1}{c|}{\textit{1.7}}& \multicolumn{1}{c|}{\textit{1.63}}& \multicolumn{1}{c|}{\textit{1.63}}& \multicolumn{1}{c|}{1.79}& \multicolumn{1}{c|}{\textit{0.0001}} \\ \hline
\multicolumn{1}{|c|}{\textbf{M-U-LSTMD}} & \multicolumn{1}{c|}{714}& \multicolumn{1}{c|}{518}& \multicolumn{1}{c|}{706}& \multicolumn{1}{c|}{836}& \multicolumn{1}{c|}{4.92}& \multicolumn{1}{c|}{2.3}& \multicolumn{1}{c|}{5.35}& \multicolumn{1}{c|}{\textit{1.49}}& \multicolumn{1}{c|}{\textit{0.0001}} \\ \hline \hline
\multicolumn{1}{|c|}{\textbf{M-B-LSTM}} & \multicolumn{1}{c|}{1441}& \multicolumn{1}{c|}{\textit{1323}}& \multicolumn{1}{c|}{\textit{1603}}& \multicolumn{1}{c|}{1602}& \multicolumn{1}{c|}{\textit{2.29}}& \multicolumn{1}{c|}{\textit{1.73}}& \multicolumn{1}{c|}{11.81}& \multicolumn{1}{c|}{5.45}& \multicolumn{1}{c|}{0.0002} \\ \hline
\multicolumn{1}{|c|}{\textbf{M-B-HBNN}} & \multicolumn{1}{c|}{2123}& \multicolumn{1}{c|}{3210}& \multicolumn{1}{c|}{3230}& \multicolumn{1}{c|}{2281}& \multicolumn{1}{c|}{2.54}& \multicolumn{1}{c|}{2.3}& \multicolumn{1}{c|}{2.6}& \multicolumn{1}{c|}{\textit{2.78}}& \multicolumn{1}{c|}{\textit{0.0001}} \\ \hline
\multicolumn{1}{|c|}{\textbf{M-B-LSTMD}} & \multicolumn{1}{c|}{\textit{954}}& \multicolumn{1}{c|}{1559}& \multicolumn{1}{c|}{1477}& \multicolumn{1}{c|}{\textit{1592}}& \multicolumn{1}{c|}{3.41}& \multicolumn{1}{c|}{3.0}& \multicolumn{1}{c|}{\textit{2.09}}& \multicolumn{1}{c|}{3.33}& \multicolumn{1}{c|}{\textit{0.0001}} \\ \hline
\end{tabular}%
\label{tab:runtime}
\end{center}
\end{table*}

The applicability of DL models to real-world scenarios strongly depends on the time necessary for training and deploying the model in a cloud resource management setup.
The three critical aspects are, therefore, the training time, the fine-tuning time (i.e. how often we update the network weights with newly available data) and the inference time. The training time depends mainly on the size of the training set, the size of the network, and the learning algorithm; the fine-tuning time depends on how often we update the weights of the deep learning model, and the inference time is related to the prediction time once the model is trained. We fix the size of the network (by hyperparameterization) and the learning algorithm, and we measure training time by varying the size of the training set in 20\%, 40\%, 60\% and 80\%. We measure fine-tuning time by varying the number of steps among 6, 12, 18 and 24, which correspond to 30, 60, 90 and 120 minutes frequency. We measure the inference time of predicting one sample. The results are computed as an average of 10 runs for all the model combinations, and we report values in Table \ref{tab:runtime}. As we can see, the models take more or less the same time for the fine-tuning and inference steps, with the HBNN often being the fastest network. The training time, instead, varies with the type of training and the prediction task. The HBNN is the slowest model (because the Bayesian layer is more complex than dense layers), except for the univariate version trained with multiple datasets. However, the training phase generally is infrequent and done offline, e.g. overnight, so it is not a critical factor. At the same time, fine-tuning and inference happen more frequently in resource management operations. We would also like to underline that the results for the S-U versions refer to the training of one single (dataset, resource) pair, which means that we need to run this phase 24 times (12 clusters $\times$ 2 resources). This also applies to the S-B (12 times) and M-U versions (2 times). We conclude that all the models can be practically deployed in real-world scenarios, with the advantage of having a model trained with multiple datasets, which does not require any parallelization of the systems for each possible cluster cell.

To summarize, results show an overall positive impact of uncertainty-aware models over the LSTM baseline, with different trade-offs given by training univariate and multivariate models on single and multiple datasets.

\subsection{On transfer learning capabilities}



Similarly to the previous section, we present and analyse the results obtained from the six prediction scenarios depicted in section \ref{sec:TrainingStrat}, where we probe the transfer learning capabilities of the M-B-HBNN model. We chose this model because it shows an efficient trade-off between performance (i.e.\ close to the best performance in terms of accuracy and service level metrics) and complexity of deployment for the experiments (i.e.\ expensive to train with multiple datasets but one single model that provides predictions for processing units and memory for all datasets). We leave for future work a more comprehensive investigation that also involves the other competitive models, such as some of the LSTMD-based.

\begin{table}[htbp]
\centering
\caption{Accuracy results for M-B-HBNN trained on scenarios \textit{All}, \textit{All FT}, \textit{All-but-one} and \textit{All-but-one FT}. In bold, the best performance. } 
\begin{tabular}{c|cc|cc|}
\cline{2-5}
 & \multicolumn{2}{c||}{\textbf{Processing units}} & \multicolumn{2}{c|}{\textbf{Memory}} \\ \hline
\multicolumn{1}{|c|}{\textbf{Scenario}} & \multicolumn{1}{c|}{\textbf{MSE}} & \multicolumn{1}{c||}{\textbf{MAE}} & \multicolumn{1}{c|}{\textbf{MSE}} & \textbf{MAE} \\ \hline
\multicolumn{1}{|c|}{\textbf{All}} & \multicolumn{1}{c|}{\textbf{0.0042}} & \multicolumn{1}{c||}{\textbf{0.0471}} & \multicolumn{1}{c|}{\textbf{0.0043}} & \textbf{0.0436} \\ \hline
\multicolumn{1}{|c|}{\textbf{All FT}} & \multicolumn{1}{c|}{0.0058} & \multicolumn{1}{c||}{0.0561} & \multicolumn{1}{c|}{0.0062} & 0.0058 \\ \hline
\multicolumn{1}{|c|}{\textbf{All-but-one}} & \multicolumn{1}{c|}{0.0044} & \multicolumn{1}{c||}{0.0481} & \multicolumn{1}{c|}{0.0045} & 0.0448 \\ \hline
\multicolumn{1}{|c|}{\textbf{All-but-one FT}} & \multicolumn{1}{c|}{0.0057} & \multicolumn{1}{c||}{0.0563} & \multicolumn{1}{c|}{0.0058} & 0.0538 \\ \hline
\end{tabular}%
\label{tab:mseddall}
\end{table}

In Table \ref{tab:mseddall}, we report the accuracy results for \textit{All}, \textit{All FT}, \textit{All-but-one} and \textit{All-but-one FT}. The model trained under \textit{All} scenario achieves the best accuracy, showing the benefit of training with data that overlap with the target domain (we can consider this scenario as a baseline for these experiments). This shows that including traces from the same provider benefits the model's accuracy. The model trained under \textit{All-but-one} achieves performance close to the one in \textit{All}, proving that knowledge can successfully transfer to the target domains. Fine-tuning instead (both in \textit{All} and \textit{All-but-one}) makes the accuracy worse; this is likely due to the fact that the model overfits the target dataset or that it makes it forget patterns learned on other traces.

\begin{table}[htbp]
\centering
\caption{Accuracy results for M-B-HBNN trained on scenario \textit{TL-GC19}. In bold, the best performance.}
\begin{tabular}{c|cc|cc|}
\cline{2-5}
 & \multicolumn{2}{c||}{\textbf{Processing units}} & \multicolumn{2}{c|}{\textbf{Memory}} \\ \hline
\multicolumn{1}{|c|}{\textbf{Scenario}} & \multicolumn{1}{c|}{\textbf{MSE}} & \multicolumn{1}{c||}{\textbf{MAE}} & \multicolumn{1}{c|}{\textbf{MSE}} & \textbf{MAE} \\ \hline
\multicolumn{1}{|c|}{\textbf{TL-GC19 All}} & \multicolumn{1}{c|}{\textbf{0.0043}} & \multicolumn{1}{c||}{\textbf{0.0474}} & \multicolumn{1}{c|}{\textbf{0.0044}} & \textbf{0.0444} \\ \hline
\multicolumn{1}{|c|}{\textbf{TL-GC19 All FT}} & \multicolumn{1}{c|}{0.0059} & \multicolumn{1}{c||}{0.0562} & \multicolumn{1}{c|}{0.0063} & 0.0546 \\ \hline
\multicolumn{1}{|c|}{\textbf{TL-GC19 All-but-one}} & \multicolumn{1}{c|}{0.0087} & \multicolumn{1}{c||}{0.0673} & \multicolumn{1}{c|}{0.0091} & 0.0681 \\ \hline
\multicolumn{1}{|c|}{\textbf{TL-GC19 All-but-one FT}} & \multicolumn{1}{c|}{0.0097} & \multicolumn{1}{c||}{0.0733} & \multicolumn{1}{c|}{0.0104} & 0.0701 \\ \hline
\end{tabular}%
\label{tab:mseddgc19}
\end{table}

We report the accuracy results of scenario TL-GC19 (where the source and target dataset are drawn from GC19) in Table \ref{tab:mseddgc19}. The table shows that the baseline \textit{TL-GC19 All} achieves the best performance. Error worsens (it doubles for MSE and increases by around 40\% for MAE in both resources ) when the source and target domains do not overlap, i.e.\ \textit{TL-GC19 All-but-one}. Thus, although the source and the target are from the same cloud provider, M-B-HBNN struggles to transfer domain knowledge effectively. One possible explanation is that the source dataset (i.e.\ seven datasets) is too small to capture the distribution-independent properties of the CG19 provider.

\begin{table}[htbp]
\begin{center}
\caption{Accuracy results for M-B-HBNN trained on scenario \textit{TL-CG19vsOther}.}



\begin{tabular}{cc|cc|cc|}
\cline{3-6}
 &  & \multicolumn{2}{c|}{\textbf{Processing units}} & \multicolumn{2}{c|}{\textbf{Memory}} \\ \hline
\multicolumn{1}{|c|}{\textbf{Target}} & \textbf{FT} & \multicolumn{1}{c|}{\textbf{MSE}} & \textbf{MAE} & \multicolumn{1}{c|}{\textbf{MSE}} & \textbf{MAE} \\ \hline
\multicolumn{1}{|c|}{\multirow{2}{*}{\textbf{GC11}}} & \xmark & \multicolumn{1}{c|}{0.1443} & 0.3176 & \multicolumn{1}{c|}{0.2467} & 0.471 \\ \cline{2-6} 
\multicolumn{1}{|c|}{} & \cmark & \multicolumn{1}{c|}{0.0588} & 0.1669 & \multicolumn{1}{c|}{0.0421} & 0.1634 \\ \hline
\multicolumn{1}{|c|}{\multirow{2}{*}{\textbf{AC18}}} & \xmark & \multicolumn{1}{c|}{0.0737} & 0.2356 & \multicolumn{1}{c|}{0.0598} & 0.2242 \\ \cline{2-6} 
\multicolumn{1}{|c|}{} & \cmark & \multicolumn{1}{c|}{0.0271} & 0.1114 & \multicolumn{1}{c|}{0.01} & 0.0614  \\ \hline
\multicolumn{1}{|c|}{\multirow{2}{*}{\textbf{AC20}}} & \xmark & \multicolumn{1}{c|}{0.3206} & 0.4658 & \multicolumn{1}{c|}{0.09} & 0.2466  \\ \cline{2-6} 
\multicolumn{1}{|c|}{} & \cmark & \multicolumn{1}{c|}{0.0522} & 0.1791 & \multicolumn{1}{c|}{0.0608} & 0.1934  \\ \hline
\end{tabular}%

\label{tab:mseddother}
\end{center}
\end{table}

\begin{table*}[htbp]
\caption{Service level metrics results of M-B-HBNN for scenarios \textit{All}, \textit{All FT}, \textit{All-but-one} and \textit{All-but-one FT}. In bold, the model with the best SR for each confidence level.}
\begin{center}
\begin{tabular}{cc|cccc||cccc|}
\cline{3-10}
 &  & \multicolumn{4}{c||}{\multirow{2}{*}{\textbf{Processing units}}} & \multicolumn{4}{c|}{\multirow{2}{*}{\textbf{Memory}}} \\
 &  & \multicolumn{4}{c||}{} & \multicolumn{4}{c|}{} \\ \hline
\multicolumn{1}{|c|}{\multirow{2}{*}{\textbf{Confidence Level}}} & \multirow{2}{*}{\textbf{Model}} & \multicolumn{1}{c|}{\multirow{2}{*}{\textbf{SR (\%)}}} & \multicolumn{1}{c|}{\multirow{2}{*}{\textbf{OP (\%)}}} & \multicolumn{1}{c|}{\multirow{2}{*}{\textbf{UP (\%)}}} & \multirow{2}{*}{\textbf{TPR (\%)}} & \multicolumn{1}{c|}{\multirow{2}{*}{\textbf{SR (\%)}}} & \multicolumn{1}{c|}{\multirow{2}{*}{\textbf{OP (\%)}}} & \multicolumn{1}{c|}{\multirow{2}{*}{\textbf{UP (\%)}}} & \multirow{2}{*}{\textbf{TPR (\%)}} \\ 
\multicolumn{1}{|c|}{} &  & \multicolumn{1}{c|}{} & \multicolumn{1}{c|}{} & \multicolumn{1}{c|}{} &  & \multicolumn{1}{c|}{} & \multicolumn{1}{c|}{} & \multicolumn{1}{c|}{} &  \\ \hhline{==========}
\multicolumn{1}{|c|}{\multirow{4}{*}{95\%}} & All & \multicolumn{1}{c|}{\textbf{94.58}} & \multicolumn{1}{c|}{20.53} & \multicolumn{1}{c|}{0.51} & 120.02  & \multicolumn{1}{c|}{95.74} & \multicolumn{1}{c|}{16.57} & \multicolumn{1}{c|}{0.25} & 116.31\\  \cline{2-10}
\multicolumn{1}{|c|}{} & All FT & \multicolumn{1}{c|}{87.47} & \multicolumn{1}{c|}{16.63} & \multicolumn{1}{c|}{1.16} & 115.47  & \multicolumn{1}{c|}{85.87} & \multicolumn{1}{c|}{9.74} & \multicolumn{1}{c|}{0.83} & 108.92\\  \cline{2-10} 
\multicolumn{1}{|c|}{} & All-but-one & \multicolumn{1}{c|}{95.7} & \multicolumn{1}{c|}{21.58} & \multicolumn{1}{c|}{0.38} & 121.2  & \multicolumn{1}{c|}{97.39} & \multicolumn{1}{c|}{19.47} & \multicolumn{1}{c|}{0.16} & 119.32\\  \cline{2-10} 
\multicolumn{1}{|c|}{} & All-but-one FT & \multicolumn{1}{c|}{94.1} & \multicolumn{1}{c|}{20.33} & \multicolumn{1}{c|}{0.47} & 119.86  & \multicolumn{1}{c|}{\textbf{95.14}} & \multicolumn{1}{c|}{16.94} & \multicolumn{1}{c|}{0.28} & 116.65\\    \hline \hline
\multicolumn{1}{|c|}{\multirow{4}{*}{97\%}} & All & \multicolumn{1}{c|}{95.95} & \multicolumn{1}{c|}{22.9} & \multicolumn{1}{c|}{0.39} & 122.5  & \multicolumn{1}{c|}{\textbf{97.11}} & \multicolumn{1}{c|}{18.53} & \multicolumn{1}{c|}{0.19} & 118.35\\   \cline{2-10}
\multicolumn{1}{|c|}{} & All FT & \multicolumn{1}{c|}{91.01} & \multicolumn{1}{c|}{19.61} & \multicolumn{1}{c|}{0.84} & 118.77  & \multicolumn{1}{c|}{90.85} & \multicolumn{1}{c|}{11.94} & \multicolumn{1}{c|}{0.55} & 111.39\\   \cline{2-10} 
\multicolumn{1}{|c|}{} & All-but-one & \multicolumn{1}{c|}{\textbf{96.75}} & \multicolumn{1}{c|}{24.18} & \multicolumn{1}{c|}{0.28} & 123.9  & \multicolumn{1}{c|}{98.03} & \multicolumn{1}{c|}{21.8} & \multicolumn{1}{c|}{0.11} & 121.69\\  \cline{2-10} 
\multicolumn{1}{|c|}{} & All-but-one FT & \multicolumn{1}{c|}{95.46} & \multicolumn{1}{c|}{22.59} & \multicolumn{1}{c|}{0.35} & 122.24  & \multicolumn{1}{c|}{96.15} & \multicolumn{1}{c|}{18.63} & \multicolumn{1}{c|}{0.21} & 118.41\\     \hline \hline
\multicolumn{1}{|c|}{\multirow{4}{*}{99\%}} & All & \multicolumn{1}{c|}{97.19} & \multicolumn{1}{c|}{27.44} & \multicolumn{1}{c|}{0.24} & 127.19  & \multicolumn{1}{c|}{98.35} & \multicolumn{1}{c|}{22.29} & \multicolumn{1}{c|}{0.11} & 122.18\\   \cline{2-10}
\multicolumn{1}{|c|}{} & All FT & \multicolumn{1}{c|}{94.5} & \multicolumn{1}{c|}{25.45} & \multicolumn{1}{c|}{0.44} & 125.0  & \multicolumn{1}{c|}{96.07} & \multicolumn{1}{c|}{16.32} & \multicolumn{1}{c|}{0.26} & 116.06\\   \cline{2-10} 
\multicolumn{1}{|c|}{} & All-but-one & \multicolumn{1}{c|}{\textbf{97.95}} & \multicolumn{1}{c|}{29.15} & \multicolumn{1}{c|}{0.15} & 128.99  & \multicolumn{1}{c|}{\textbf{99.04}} & \multicolumn{1}{c|}{26.23} & \multicolumn{1}{c|}{0.06} & 126.17\\   \cline{2-10} 
\multicolumn{1}{|c|}{} & All-but-one FT & \multicolumn{1}{c|}{97.35} & \multicolumn{1}{c|}{26.94} & \multicolumn{1}{c|}{0.2} & 126.74  & \multicolumn{1}{c|}{97.71} & \multicolumn{1}{c|}{21.86} & \multicolumn{1}{c|}{0.13} & 121.73\\ \hline
\end{tabular}
\label{tab:trainingstatisticsall}
\end{center}
\end{table*}

Accuracy results of scenario TL-GC19vsOther (where GC19 is the source dataset and the target dataset is one of the other three datasets) are reported in Table \ref{tab:mseddother}. The model's error is quite high for all three target domains (in particular for AC20), which means that, again, the model struggle to generalise on unseen domains (this time from different providers). However, this time fine-tuning on the target domain helps to improve the performance (especially for AC20), although error values are an order of magnitude higher than the \textit{All} baseline.

\begin{figure}[htbp]
\centerline{\includegraphics[width=\linewidth]{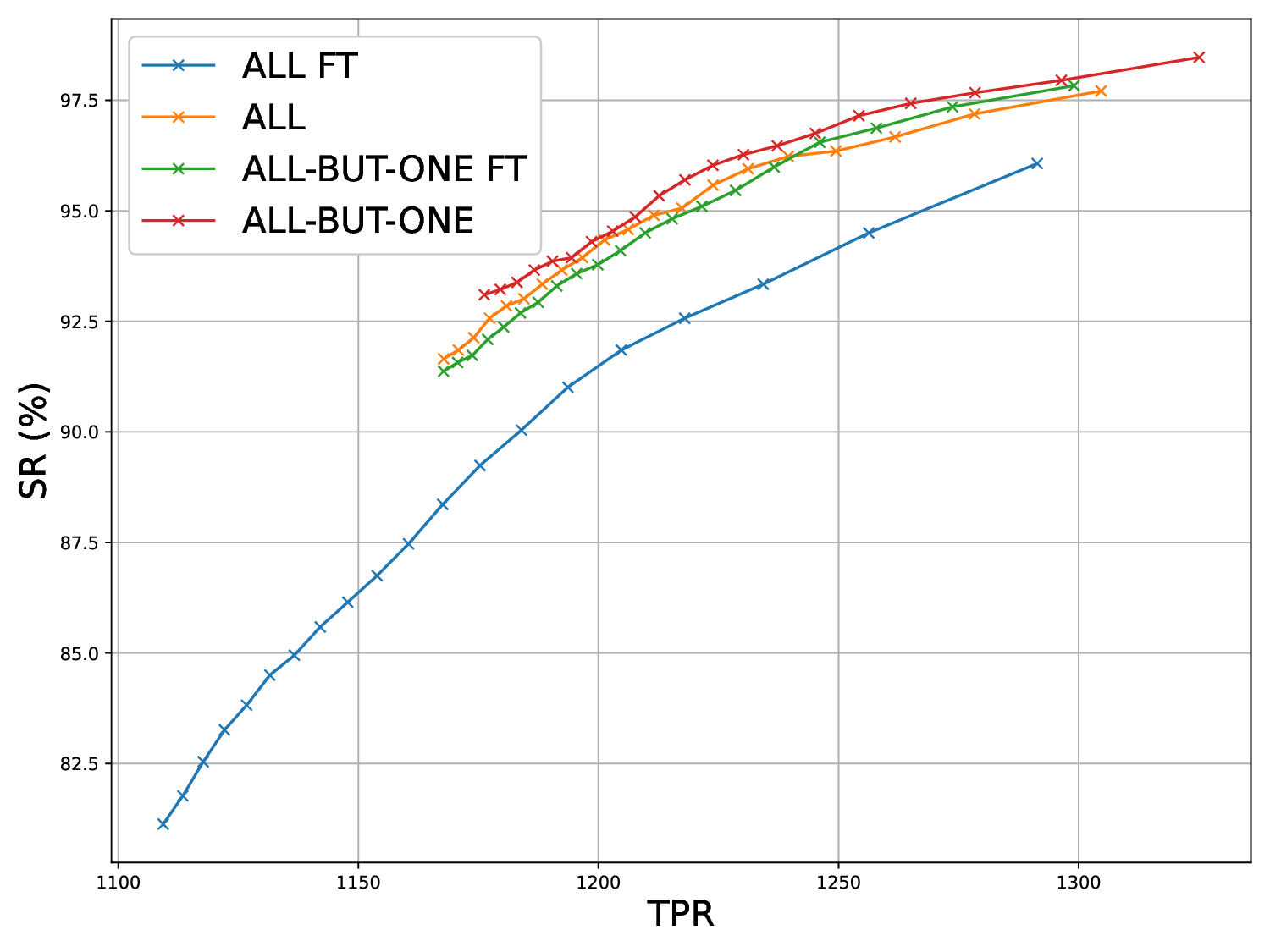}}
\caption{Total predicted processing units for \textit{All}, \textit{All FT}, \textit{All-but-one} and \textit{All-but-one FT}. The closer the curve to the top left corner, the better the model.}
\label{fig:alltprcpu}
\end{figure}


\begin{figure}[htbp]
\centerline{\includegraphics[width=\linewidth]{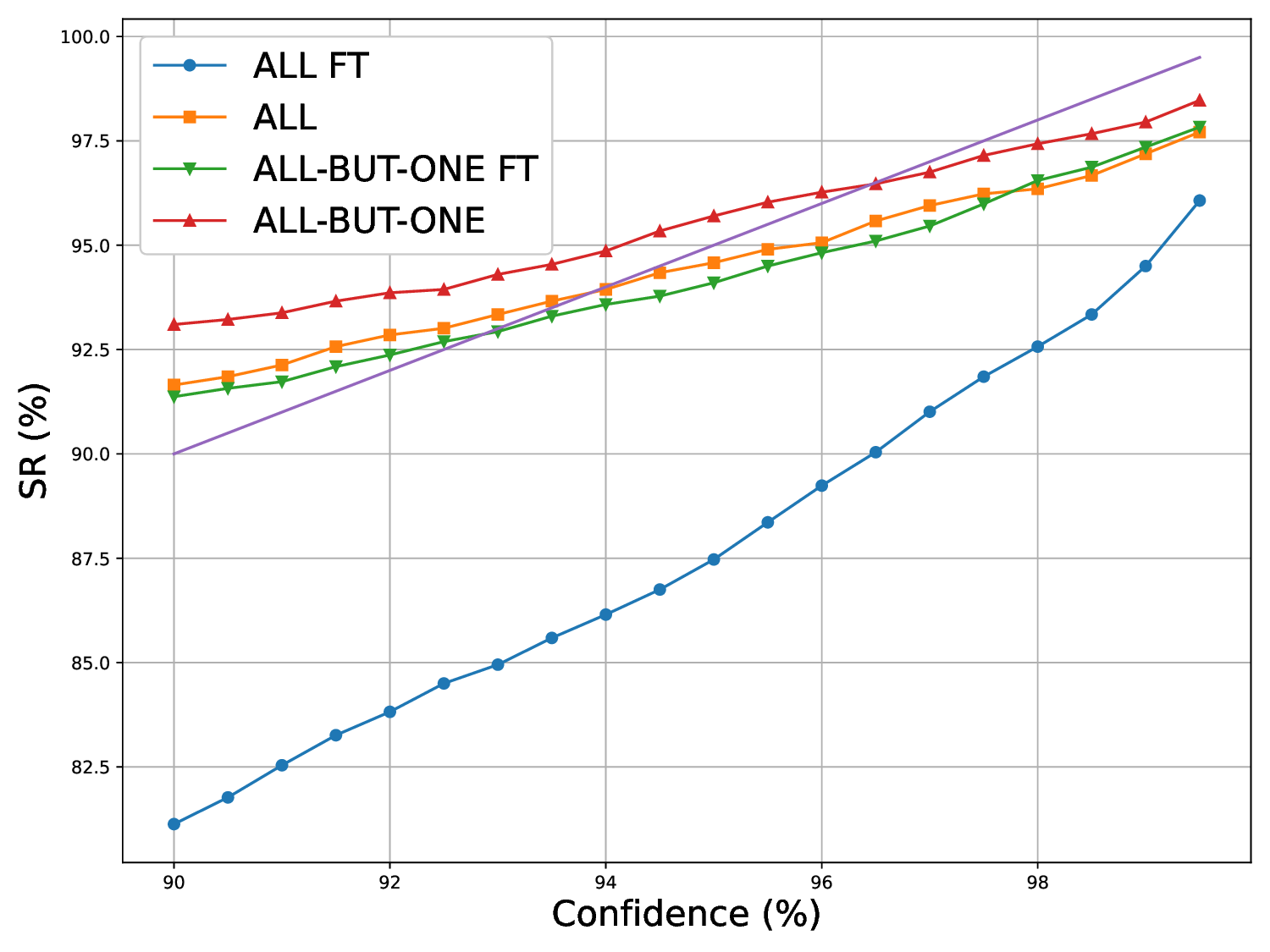}}
\caption{Target confidence level versus SR for processing units prediction for \textit{All}, \textit{All FT}, \textit{All-but-one} and \textit{All-but-one FT}. The closer the plot is to the line $y=x$, the more accurate the model under this metric.}
\label{fig:allclcpu}
\end{figure}

\begin{figure}[htbp]
\centerline{\includegraphics[width=\linewidth]{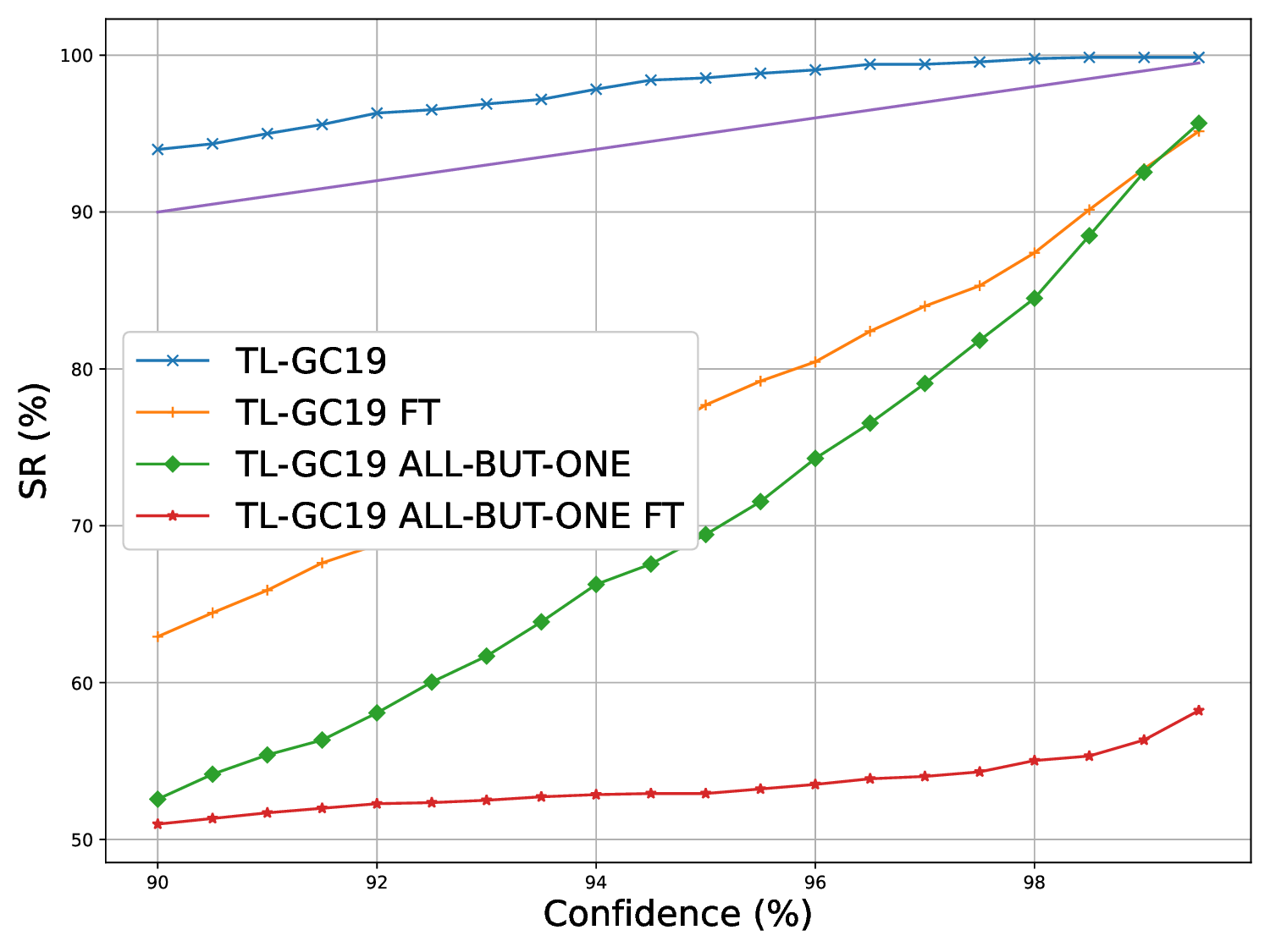}}
\caption{Target service level versus SR for processing units prediction for \textit{TL-GC19}. The closer the plot is to the line $y=x$, the more accurate the model under this metric.}
\label{fig:gc19clcpu}
\end{figure}

We also analyse the TL capabilities of M-B-HBNN regarding the service level metrics. Table \ref{tab:trainingstatisticsall}, Figures \ref{fig:alltprcpu} and \ref{fig:allclcpu} report the results for scenarios \textit{All}, \textit{All FT}, \textit{All-but-one} and \textit{All-but-one FT}. Table \ref{tab:trainingstatisticsfgc19} and Figure \ref{fig:gc19clcpu} report the results for scenario \textit{TL-GC19}. Because results for scenario \textit{TL-GC19vsOther} are poor (i.e.\ M-B-HBNN fails to generalise on unseen cloud providers), we avoid reporting and detailed analysis. From the aforementioned tables and figure, the main takeaways are that (i) \textit{All-but-one} achieves similar performance to \textit{All}, confirming TL capabilities found earlier by the accuracy metrics (ii) the same TL capabilities do not hold when the source and target domains are from GC19, i.e.\ \textit{TL-GC19}, and (iii) fine-tuning always harm the performance.

\begin{table*}[htbp]
\caption{Service level metrics results of M-B-HBNN for scenario \textit{TL-GC19}. In bold, the model with the best SR for each confidence level.}
\begin{center}
\begin{tabular}{cc|cccc||cccc|}
\cline{3-10}
 &  & \multicolumn{4}{c||}{\multirow{2}{*}{\textbf{Processing units}}} & \multicolumn{4}{c|}{\multirow{2}{*}{\textbf{Memory}}} \\
 &  & \multicolumn{4}{c||}{} & \multicolumn{4}{c|}{} \\ \hline
\multicolumn{1}{|c|}{\multirow{2}{*}{\textbf{Confidence Level}}} & \multirow{2}{*}{\textbf{Model}} & \multicolumn{1}{c|}{\multirow{2}{*}{\textbf{SR (\%)}}} & \multicolumn{1}{c|}{\multirow{2}{*}{\textbf{OP (\%)}}} & \multicolumn{1}{c|}{\multirow{2}{*}{\textbf{UP (\%)}}} & \multirow{2}{*}{\textbf{TPR (\%)}} & \multicolumn{1}{c|}{\multirow{2}{*}{\textbf{SR (\%)}}} & \multicolumn{1}{c|}{\multirow{2}{*}{\textbf{OP (\%)}}} & \multicolumn{1}{c|}{\multirow{2}{*}{\textbf{UP (\%)}}} & \multirow{2}{*}{\textbf{TPR (\%)}} \\ 
\multicolumn{1}{|c|}{} &  & \multicolumn{1}{c|}{} & \multicolumn{1}{c|}{} & \multicolumn{1}{c|}{} &  & \multicolumn{1}{c|}{} & \multicolumn{1}{c|}{} & \multicolumn{1}{c|}{} &  \\ \hhline{==========}
\multicolumn{1}{|c|}{\multirow{4}{*}{95\%}} &  TL-GC19 & \multicolumn{1}{c|}{\textbf{98.55}} & \multicolumn{1}{c|}{10.85} & \multicolumn{1}{c|}{0.04} & 112.04  & \multicolumn{1}{c|}{\textbf{97.18}} & \multicolumn{1}{c|}{12.29} & \multicolumn{1}{c|}{0.13} & 109.12\\   \cline{2-10} 
\multicolumn{1}{|c|}{} & TL-GC19 FT & \multicolumn{1}{c|}{77.7} & \multicolumn{1}{c|}{4.96} & \multicolumn{1}{c|}{0.85} & 105.34  & \multicolumn{1}{c|}{89.36} & \multicolumn{1}{c|}{7.89} & \multicolumn{1}{c|}{0.41} & 104.45\\  \cline{2-10} 
\multicolumn{1}{|c|}{} & TL-GC19 All-but-one & \multicolumn{1}{c|}{69.44} & \multicolumn{1}{c|}{4.07} & \multicolumn{1}{c|}{1.05} & 104.26  & \multicolumn{1}{c|}{74.44} & \multicolumn{1}{c|}{4.75} & \multicolumn{1}{c|}{1.12} & 100.59\\ \cline{2-10} 
\multicolumn{1}{|c|}{} & TL-GC19 All-but-one FT & \multicolumn{1}{c|}{52.93} & \multicolumn{1}{c|}{13.62} & \multicolumn{1}{c|}{22.08} & 92.78  & \multicolumn{1}{c|}{63.79} & \multicolumn{1}{c|}{46.63} & \multicolumn{1}{c|}{10.35} & 133.24\\ 
   \hline \hline
\multicolumn{1}{|c|}{\multirow{4}{*}{97\%}}  & TL-GC19 & \multicolumn{1}{c|}{\textbf{99.42}} & \multicolumn{1}{c|}{12.53} & \multicolumn{1}{c|}{0.02} & 113.75  & \multicolumn{1}{c|}{\textbf{98.19}} & \multicolumn{1}{c|}{14.07} & \multicolumn{1}{c|}{0.1} & 110.94\\  \cline{2-10} 
\multicolumn{1}{|c|}{} & TL-GC19 FT &\multicolumn{1}{c|}{84.0} & \multicolumn{1}{c|}{6.04} & \multicolumn{1}{c|}{0.59} & 106.69  & \multicolumn{1}{c|}{92.9} & \multicolumn{1}{c|}{9.25} & \multicolumn{1}{c|}{0.28} & 105.94\\  \cline{2-10} 
\multicolumn{1}{|c|}{} & TL-GC19 All but-one &\multicolumn{1}{c|}{79.07} & \multicolumn{1}{c|}{5.22} & \multicolumn{1}{c|}{0.65} & 105.81  & \multicolumn{1}{c|}{83.06} & \multicolumn{1}{c|}{6.35} & \multicolumn{1}{c|}{0.7} & 102.61\\\cline{2-10} 
\multicolumn{1}{|c|}{} & TL-GC19 All-but-one FT & \multicolumn{1}{c|}{54.02} & \multicolumn{1}{c|}{14.45} & \multicolumn{1}{c|}{20.67} & 95.01  & \multicolumn{1}{c|}{65.03} & \multicolumn{1}{c|}{48.93} & \multicolumn{1}{c|}{9.27} & 136.63\\ 
  \hline \hline
\multicolumn{1}{|c|}{\multirow{4}{*}{99\%}} & TL-GC19 & \multicolumn{1}{c|}{\textbf{99.86}} & \multicolumn{1}{c|}{15.74} & \multicolumn{1}{c|}{0.01} & 116.96  & \multicolumn{1}{c|}{\textbf{98.84}} & \multicolumn{1}{c|}{17.47} & \multicolumn{1}{c|}{0.06} & 114.38\\  \cline{2-10} 
\multicolumn{1}{|c|}{} & TL-GC19 FT & \multicolumn{1}{c|}{92.76} & \multicolumn{1}{c|}{8.28} & \multicolumn{1}{c|}{0.29} & 109.23  & \multicolumn{1}{c|}{96.6} & \multicolumn{1}{c|}{11.93} & \multicolumn{1}{c|}{0.14} & 108.76\\  \cline{2-10} 
\multicolumn{1}{|c|}{} & TL-GC19 All-but-one & \multicolumn{1}{c|}{92.54} & \multicolumn{1}{c|}{7.74} & \multicolumn{1}{c|}{0.23} & 108.74  & \multicolumn{1}{c|}{92.9} & \multicolumn{1}{c|}{9.7} & \multicolumn{1}{c|}{0.26} & 106.41\\\cline{2-10} 
\multicolumn{1}{|c|}{} & TL-GC19 All-but-one FT & \multicolumn{1}{c|}{56.34} & \multicolumn{1}{c|}{16.1} & \multicolumn{1}{c|}{18.1} & 99.23  & \multicolumn{1}{c|}{69.08} & \multicolumn{1}{c|}{53.47} & \multicolumn{1}{c|}{7.4} & 143.03\\   
  \hline
\end{tabular}
\label{tab:trainingstatisticsfgc19}
\end{center}
\end{table*}

\begin{figure}[htbp]
\centerline{\includegraphics[width=\linewidth]{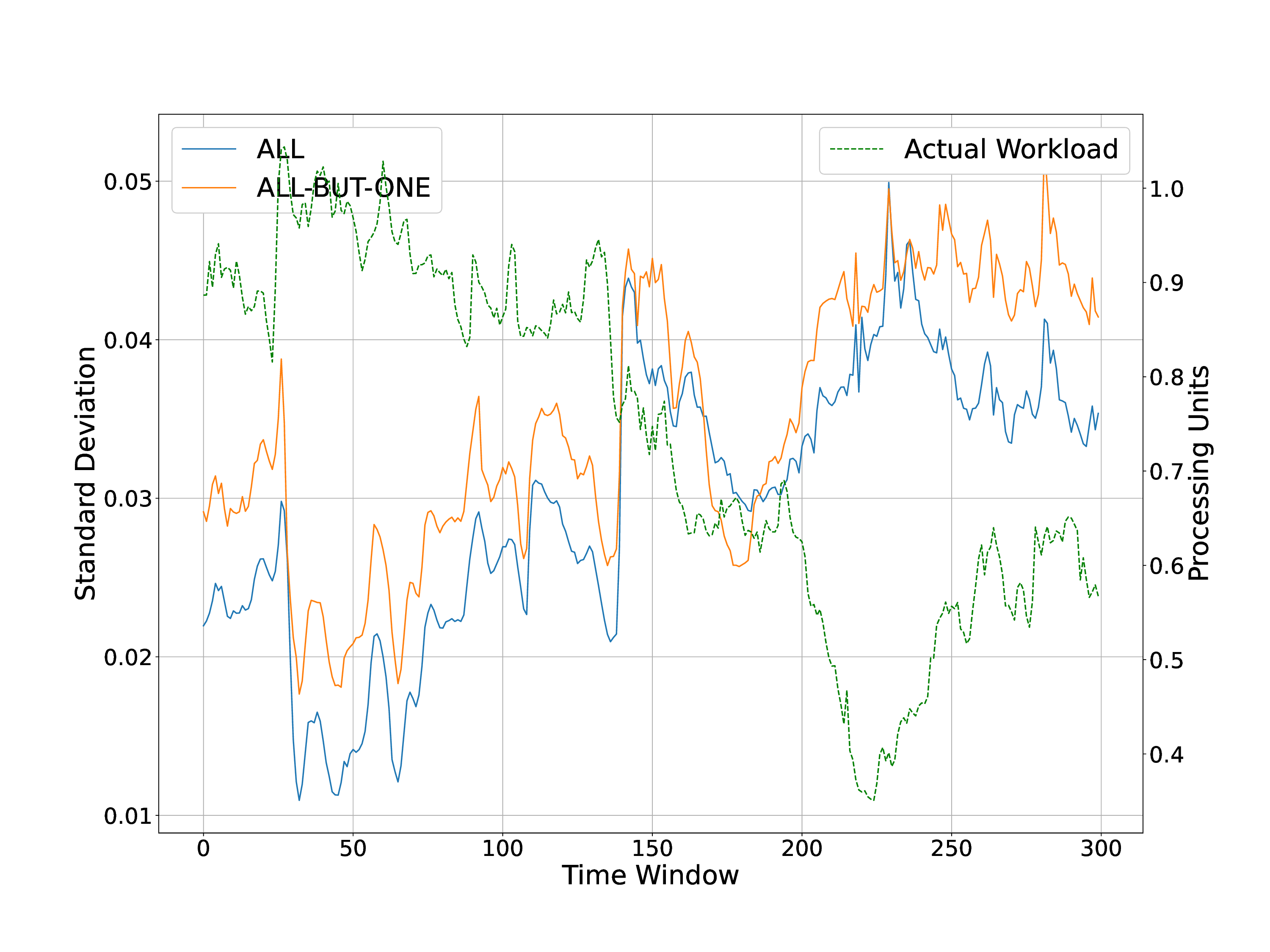}}
\caption{Standard deviation variability for across inference on the test set \textit{All} and \textit{All-but-one}. Uncertainty for All-but-one is higher for predictions on unseen data.}
\label{fig:alluncertainty}
\end{figure}

Finally, we analyse how uncertainty varies when inference is made on different target datasets. In Fig. \ref{fig:alluncertainty}, we plot the values of the standard deviation (that governs the size of the confidence interval, and its magnitude is proportional to the amount of uncertainty in the prediction) of \textit{All} vs \textit{All-but-one}. The plot shows that the standard deviation values for \textit{All-but-one} are higher than the values for \textit{All}. This is because when the model predicts unseen datasets (missing from the training) the model is more uncertain about its predictions. This further confirms the usefulness of uncertainty as an additional feature to predictions.

Overall, the results of our TL experiments show that M-B-HBNN can transfer domain knowledge effectively only when the source domain used for training is sufficiently big and sufficiently similar to the target domain. On the contrary, it struggles to generalise well when the source domain is small or the target domain is very different from the target domain.


\section{Conclusion and Future Work}\label{sec:conclusion}

Uncertainty-aware workload forecasting provides richer information than a simple pointwise prediction, and this can help cloud resource managers to optimize their decision-making process. In this paper, we investigated the performance of probabilistic models that can capture the aleatoric and epistemic uncertainties of their predictions by forecasting a probability distribution of the future workload demand. 

We first evaluated univariate and bivariate versions of HBNN and LSTMD that can predict processing units and memory usage of workload traces (i.e.\ datasets) from Google Cloud and Alibaba. Experiments show that training one model on multiple datasets benefits the learning task more than training many models on single datasets. While evaluation through traditional pointwise accuracy metrics does not provide a clear picture of the effectiveness of these models, a prediction and its uncertainty based on confidence levels allow one to tailor corresponding desired service levels and therefore have a positive impact on cloud service metrics. Also, the runtime of all models is efficient enough that they can be deployed in real-world applications. In particular, we found that the bivariate HBNN trained on multiple datasets (i.e.\ M-B-HBNN) provide a good trade-off between performance and complexity of deployment for the second part of our investigation.

Indeed, we further probed the transfer learning capabilities of uncertainty-aware models, as models' generalisation on out-of-distribution target domains is under-investigated in workload forecasting. In practice, we checked whether the M-B-HBNN, trained in zero-shot or fine-tuning scenarios, can transfer knowledge across related (to different extents) domains. Experiments reveal that the more source and target domains are different (particularly when the source and target are from different cloud providers), the more performance degrades, and fine-tuning most of the time does not help the knowledge transfer. Acceptable transfer learning performances can be achieved when the training source domain is big enough to account for the target domain diversity from the source domain. However, we still believe cloud providers should explore ways of quickly building forecasting models in scenarios where new (unseen) or little data is available. Therefore, further investigation into transfer learning is needed, particularly on the relationship between the source and target domain, to make the transfer successful.

In future works, we plan to tailor the training to a desired confidence level by explicitly, for example, leveraging a non-symmetrical loss function. Then, we will design and implement a fully Bayesian neural network to capture the full posterior distribution over the network weights and outputs that hopefully makes uncertainty estimation more robust and effective. Also, we would like to explore the effectiveness of uncertainty-aware models for multi-step-ahead workload prediction. Another option is to keep the framework we developed in this paper but focus on machine-level workload forecasting as a different scope to assess the usefulness of uncertainty-aware predictions. Finally, exploring how uncertainty-aware predictions can be exploited for scheduling, resource allocation, scaling and balancing problems in a cloud computing environment is of extreme interest to practitioners. Thus, there is work to do on integrating the uncertainty-aware models and the resource allocators (optimizers) in the same pipeline.

\ifCLASSOPTIONcompsoc
  \section*{Acknowledgement}
\else
  \section*{Acknowledgement}
\fi

This work was conducted with the financial support of Science Foundation Ireland under Grant Nos. 18/CRT/6223, 16/RC/3918 and 12/RC/2289-P2. which are co-funded under the European Regional Development Fund; by the EU Horizon projects Glaciation (GA No. 101070141) and TAILOR (GA No. 952215), and by the Google Cloud Research Credits program with the award GCP203677602. Andrea Rossi is the corresponding author.

\section*{Funding information}

Science Foundation Ireland 18/CRT/6223

Science Foundation Ireland 16/RC/3918

Science Foundation Ireland 12/RC/2289-P2

EU Horizon Glaciation 101070141

EU Horizon TAILOR (GA No. 952215)

Google Cloud Research Credits program GCP203677602

\section*{conflict of interest}
The authors declare no conflicts of interest.

\section*{data availability statement}
The data used in the experiments are available on the GitHub repository \url{https://github.com/andreareds/TowardsUncertaintyAwareWorkloadPrediction}

\ifCLASSOPTIONcaptionsoff
  \newpage
\fi



%

\bibliographystyle{IEEEtranN}
\bibliography{IEEEabrv, references}

\end{document}